\documentclass[12pt, a4paper]{article}
\usepackage{jheppub}
\usepackage[utf8]{inputenc}
\usepackage[dvipsnames]{xcolor}
\usepackage[normalem]{ulem}
\usepackage[bb=boondox]{mathalfa}
\usepackage[export]{adjustbox}
\usepackage{mathtools,resizegather,amsmath,amsfonts,amssymb,mathrsfs,amsthm,amsbsy,multirow,multicol,subcaption,dcolumn,makecell,scalerel,verbatim,braket,slashed, graphicx, physics,bbm,bm,algpseudocode, algorithm, upgreek, xfrac}

\def\be{\begin{equation}}
\def\ee{\end{equation}}

\preprint{USTC-ICTS/PCFT-26-38\\ \hspace*{10.3cm} 
LITP-26-11 \\
\hspace*{10.3cm} 
IFT-UAM/CSIC-26-81}

\title{\Large Low-temperature Quantum-corrected Holographic Transport with Momentum Relaxation}

\author[a]{Suman Das\,\href{https://orcid.org/0000-0002-0053-3187}
{\includegraphics[scale=0.05]{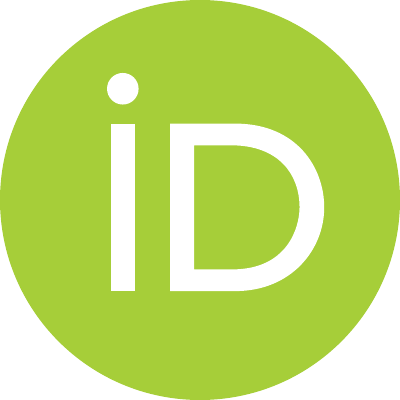}}}

\author[b, c]{,\, Sabyasachi Maulik\,\href{https://orcid.org/0000-0003-4132-2666}
{\includegraphics[scale=0.05]{orcidid.pdf}}}
\author[d]{,\, Leopoldo A. Pando Zayas\,\href{https://orcid.org/0000-0003-0727-5958}{\includegraphics[scale=0.05]{orcidid.pdf}}}

\author[d]{,\, and Jingchao Zhang\,\href{https://orcid.org/0009-0009-3296-6495}
{\includegraphics[scale=0.05]{orcidid.pdf}}}

\emailAdd{suman.das@ift.csic.es}
\emailAdd{mauliks@ustc.edu.cn}
\emailAdd{lpandoz@umich.edu}
\emailAdd{jingchaz@umich.edu}

\affiliation[a]{Instituto de F\'{i}sica Te\'{o}rica UAM/CSIC, Calle Nicol\'{a}s Cabrera 13-15, Madrid, E-28049, Spain.}

\affiliation[b]{Interdisciplinary Center for Theoretical Study, University of Science and Technology of China, Hefei, Anhui 230026, China.}
\affiliation[c]{Peng Huanwu Center for Fundamental Theory, Hefei, Anhui 230026, China.}
\affiliation[d]{Leinweber Institute for Theoretical Physics, University of Michigan, Ann Arbor, MI 48109, USA.}

\abstract{We determine the quantum corrections to transport arising from fluctuations of the near-AdS${}_2$ throat of near-extremal black branes in holographic models with momentum relaxation. By computing the shear viscosity and electrical conductivity at both zero and finite chemical potential, we uncover a universal low-temperature enhancement of transport generated by Schwarzian quantum fluctuations. Specifically, transport coefficients extracted from retarded Green's functions increase throughout the regime $C T \ll 1$. For operators with Schwarzian scaling dimension $\Delta >1$, this enhancement is preceded by a universal minimum at a characteristic temperature $T_{\rm min} \propto C^{-1}$, leading to a non-monotonic temperature dependence strikingly similar to that observed in many correlated materials. In contrast, for the case $\Delta =1$, relevant for the electrical conductivity, the transport coefficient evolves monotonically toward a constant value. Our results identify universal signatures of near-horizon quantum gravity in the transport properties of holographic quantum matter.}

\begin{document}

\maketitle
\flushbottom

\section{Introduction}\label{sec:intro}

The AdS/CFT correspondence \cite{Maldacena:1997re, Witten:1998qj, Gubser:1998bc} provides a remarkable connection between gravity in  asymptotically AdS spacetimes and a dual conformal field theory via an identification of generating functions,
\begin{equation}
    Z_{\text{QG}}\!\left[\phi_{(0)}\right]
    \;=\;
    Z_{\text{CFT}}\!\left[J\right]\Big|_{J=\phi_{(0)}} ,
\end{equation}
where $\phi_{(0)}$ denotes the boundary values of bulk fields that act as sources $J$ for the corresponding operators in the CFT. Such correspondence opens a new window  into the dynamics of strongly coupled quantum field theories that lie beyond the reach of conventional perturbative techniques. According to the correspondence, a strongly interacting large-$N$ quantum field theory at
strong \textquotesingle t~Hooft coupling is dual to a (semi)classical gravitational theory in an asymptotically AdS spacetime, where computations are often technically tractable. Using this holographic framework, valuable insights have been obtained into a wide class of strongly correlated systems, including phenomena such as holographic superconductors, strange metals, and non-Fermi-liquid phases, among others \cite{Gubser:2008px, Hartnoll:2008vx, Hartnoll:2008kx, Lee:2008xf, Liu:2009dm, Cubrovic:2009ye, Lee:2010ii, Lee:2010qs, Zaanen:2015oix, Baggioli:2019rrs}.

One particularly useful and interesting regime to consider in such strongly interacting systems is the hydrodynamic regime, and AdS/CFT has proven to be highly successful in its description
\cite{Policastro:2002se, Bhattacharyya:2007vjd, Rangamani:2009xk, Bredberg:2011jq, Hubeny:2011hd}. In quantum field theory, the hydrodynamic regime describes the universal long-wavelength, low-frequency limit in which the dynamics is governed solely by conservation laws. In this regime, microscopic details enter only through a finite set of transport coefficients, which quantify the dissipative response of the system to external perturbations. These coefficients control late-time relaxation and collective excitations and provide direct probes of the strength of interactions.

For strongly coupled gauge theories, the computation of transport coefficients such as the shear viscosity, $\eta $, or the conductivity, $\sigma$, is in general very difficult. The AdS/CFT correspondence turns out to be of crucial utility in such cases, as it allows these quantities to be computed in a remarkably simpler way in the dual gravitational description. A well-known example of strong coupling universality is the ratio of shear viscosity to entropy density \cite{Policastro:2001yc, Buchel:2003tz}. Broad studies of this ratio in holographic theories gave impetus to the Kovtun--Son--Starinets (KSS) bound \cite{Kovtun:2003wp, Kovtun:2004de} -- which states that
\begin{equation}
    \frac{\eta}{s}\geq\frac{1}{4\pi}.
\end{equation}
This result is robust for two-derivative Einstein gravity with translationally and rotationally invariant states. Violations can occur either by including higher-derivative corrections to Einstein gravity \cite{Kats:2007mq, Buchel:2008vz} or by introducing anisotropy in the system \cite{Rebhan:2011vd, Mamo:2012sy, Jain:2014vka, Critelli:2014kra, Jain:2015txa}.

The treatment of black holes at very low temperatures has recently been shown to received important quantum corrections \cite{Iliesiu:2020qvm, Heydeman:2020hhw, Boruch:2022tno, Turiaci:2023wrh}. The physics of this strong quantum fluctations affect the holographic dictionary as discussed in \cite{Daguerre:2023cyx,Liu:2024gxr}. In particular,  an interesting novel behavior of shear viscosity to entropy density ratio, $\eta/s$,  has been reported \cite{Nian:2025oei,PandoZayas:2025snm,Cremonini:2025yqe,Gouteraux:2025exs,Kanargias:2025vul}. All these discussions, however,  have taken place in models with translational invariance.

A difficulty with translation invariant systems is the absence of momentum relaxation and as a result the DC conductivity diverges. This is not the case in realistic  condensed matter systems, where momentum dissipates. There are several ways to incorporate momentum relaxation holographically. Among them, there is coupling the charge carriers, treated in the probe approximation, to a parametrically larger neutral sector that acts as a reservoir into which their momentum can dissipate \cite{Karch:2007pd, Hartnoll:2009ns, Charmousis:2010zz, Gouteraux:2011ce, Faulkner:2010da, Faulkner:2013bna}. One can consider breaking translational invariance explicitly by introducing impurities \cite{Hartnoll:2007ih, Hartnoll:2008hs, Lucas:2014zea}. Similarly, one can consider introducing holographic lattices \cite{Horowitz:2012ky, Horowitz:2012gs, Horowitz:2013jaa, Ling:2013nxa}, where translational symmetry is broken by a spatially inhomogeneous source for a neutral scalar field, or using Q-lattice models \cite{Donos:2013eha, Donos:2014uba} in which the global phase of a complex scalar breaks translational invariance. Other realizations include breaking bulk diffeomorphism invariance via massive gravity \cite{deRham:2010kj}, which corresponds to breaking translational invariance in the boundary theory \cite{Vegh:2013sk, Davison:2013jba, Blake:2013bqa}. The approach  we are going to focus in this manuscript considers models with massless scalar fields with spatially varying sources, which lead to a non-conserved energy-momentum tensor \cite{Andrade:2013gsa, Kim:2014bza, Taylor:2014tka, Davison:2014lua}.

{\it In this manuscript we bring aspects of low-temperature quantum corrections to bear in the computation of transport coefficients in a model of momentum relaxation. }

The rest of the article is organized as follows. In Section \ref{sec:RN-AdS_review} we review the model of momentum relaxation and consider its low-temperature limit.  In Section \ref{sec:classical_Green} we review previous results for the viscosity and conductivity in the linear axion model realizing momentum relaxation. Section \ref{sec:quantum_Green} presents the effects of incorporating quantum corrections in the corresponding transport coefficients (shear viscosity and electrical conductivity). Section \ref{Sec:ChemicalPotential} contains our main results, there we extend the discussion of transport coefficients to a situation with non-vanishing chemical potential by considering a background with electric field.  We conclude in Section \ref{sec:conclusions}. Appendix \ref{App:GreenDelta} derives an expression for the quantum-corrected Green's function for arbitrary $\Delta$ which supports the universality of certain behavior. In Appendix \ref{app:normalization-boundary-coefficients}, we clarify the precise normalization of the Green's function.

\section{Isotropic linear axion model} \label{sec:RN-AdS_review}

The linear axion model was originally introduced in \cite{Andrade:2013gsa} to study the phenomenon of momentum relaxation in the holographic framework. The model in the bulk consists of two massless scalar fields, $\phi^{j}$ minimally coupled to Einstein-Hilbert gravity with a negative cosmological constant. We focus on $\left(3+1\right)$-dimensional bulk spacetime with the action
\begin{equation}\label{action0}
    I=\frac{1}{16 \pi G} \int d^4x \, \sqrt{-g} 
    \left( R+\frac{6}{L^2} - \frac{1}{2} \sum_{j=1}^{2} \left(\partial \phi^{j} \right)^{2}  \right),
\end{equation}
where $G$ is the gravitational constant, and $L$ is the AdS$_{4}$ radius. The relevant solution to the classical equations of motion of the theory are homogeneous and well known in closed form
\begin{subequations}
\begin{align}
    ds^2 &= -f(r)\,dt^2+\frac{dr^2}{f(r)} + \frac{r^2}{L^2} 4\pi V_{2} \left( dx_1^2 + dx_2^2 \right),\\
    f(r) &= \frac{r^2}{L^2} - \frac{\alpha^2 L^4}{2}-\frac{r_+}{r} \left(\frac{r_+^2}{L^2} - \frac{\alpha^2 L^4}{2} \right).
\end{align} \label{metric}
\end{subequations}
The matter sector breaks translational symmetry along the spatial directions of the AdS$_{4}$ conformal boundary according to 
\begin{equation}
    \phi^i = \sqrt{4\pi V_{2}}\,\alpha\, x_i, \quad \text{for} \quad i \in \{1, 2\} . \label{gaugefield}
\end{equation}
For simplicity, we consider that $\left(x_{1}, x_{2} \right)$ are dimensionless coordinates of periodicity $L_{1, 2}$, and the space spanned by them is a torus of volume $4\pi V_{2}$. The complete coordinate ranges are
\[
t\in\left(-\infty,\infty\right),\quad 
r\in\left[r_+,\infty\right),\quad 
x_1\in\left[0, L_{1}\right],\quad
x_2\in\left[0, L_{2}\right].
\]
We set $L=1$ from this point onwards. Here, $r_+ > 0$ denotes the position of the outer horizon which implies taking $|\alpha| \leq \frac{\sqrt{6} r_+}{L^{3}}$. The Hawking temperature $T_H$ and energy density $\varepsilon$ associated with the black brane geometry are given by
\begin{equation} \label{eq:temp_and_energy}
    T_H=\frac{1}{8 \pi r_+} \left(6r_+^2 -\alpha^2 \right), \qquad \varepsilon=\frac{r_+}{16 \pi G}(2r_+^2-\alpha^2).
\end{equation}
We can naturally  assign a Bekenstein-Hawking entropy to the black brane, and its density is given by
\begin{equation}
    s=\frac{r_+^2}{4 G}.
\end{equation}
The geometry \eqref{metric} admits an extremal limit when $|\alpha|=\sqrt{6}\, r_+$. Following Eq. \eqref{eq:temp_and_energy} one obtains that $T_H=0$, as expected, and  
\begin{equation}\label{eq:some_relations}
    r_+ \equiv r_0 = \frac{\left|\alpha \right|}{\sqrt{6}}, \quad s_0 = \frac{\alpha^2}{24 G}, \quad \varepsilon_{0} = -\frac{\alpha^3}{24 \sqrt{6} \pi  G}
\end{equation}
It is noteworthy that the extremal energy density is negative, which is a reminiscent of AdS soliton behavior \cite{Horowitz:1998ha}.

Let us consider a bulk scalar field $\phi(r, t, x^i)$ having a spatially varying source of the form $\phi^{(0)} \propto \alpha x^i$. In this case the bulk stress tensor remains homogeneous, and the Ward identity for the boundary stress tensor becomes
\begin{equation}
    \partial_{\mu} T^{\mu \nu}=\partial^{\nu} \phi^{(0)} \langle O \rangle ,
\end{equation}
where $\langle O \rangle$ is the vacuum expectation value of the operator dual to $\phi$. This implies the non-conservation of the momentum density $\langle P^i \rangle \equiv \langle T^{t i} \rangle$, which in turn leads to momentum relaxation. Although this choice of source yields a homogeneous bulk metric, it does not automatically guarantee isotropy. Isotropy can be restored by introducing $d-1$ scalar fields in a $(d+1)$-dimensional bulk \cite{Andrade:2013gsa}. In our case, with $d=3$, two scalar fields are sufficient. 

Let us first discuss how the thermodynamic properties of the background behave as we turn on a small temperature away from extremality. We work in an ensemble where the momentum relaxation scale, $\alpha$ remains fixed at its extremal value. Using the expression for Hawking temperature from Equation \eqref{eq:temp_and_energy} we find that upon slightly perturbing the black brane away from extremality, the outer horizon as a function of temperature is given by\footnote{We write $T_{H} = T$ for simplicity.}
\begin{align}
    r_+ = r_0 \left( 1+2\pi \sqrt{\frac{2}{3}} \,  \frac{T}{\alpha}+ \frac{4\pi^2}{3} \frac{T^2}{\alpha^2} +\ldots \right),\quad \alpha = \sqrt{6} r_{0}.
\end{align}

Similarly, the energy and entropy density also receive corrections
\begin{align}
    \varepsilon &= \varepsilon_0 \left( 1-4\pi^2 \,  \frac{T^2}{\alpha^2}+ \ldots \right), \\
    s &= s_0 \left( 1+4\pi \sqrt{\frac{2}{3}} \,  \frac{T}{\alpha}+ \frac{16\pi^2}{3} \frac{T^2}{\alpha^2} +\ldots \right).
\end{align}
The linear in temperature behavior of the entropy compared to the quadratic in temperature scaling of the energy was identified in \cite{Preskill:1991tb} to signify a breakdown of semiclassical black hole thermodynamics, and motivates the need to include quantum corrections as the temperature approaches zero \cite{Iliesiu:2020qvm, Heydeman:2020hhw, Boruch:2022tno,Turiaci:2023wrh, Iliesiu:2022onk}. The energy scale below which this breakdown occurs could be obtained as
\begin{equation}
\varepsilon_{\text{gap}} \equiv 4\pi^2
\left(\frac{\partial s}{\partial T} \right)^{-1}
= \frac{24 \pi G}{\alpha}\sqrt{\frac{3}{2}}
= \frac{12 \pi G}{r_{0}}.
\end{equation}
It is well known that the near-horizon metric in the extremal limit of any black hole such as \eqref{metric} contains  an AdS$_{2}$ factor. The easiest way to verify this fact is to consider the extremal limit $T_{H} = 0$, $\alpha^{2} = 6 r_{0}^{2}$, and the near-horizon scaling
\begin{equation}
    r = r_{0} + \frac{\lambda}{3 Y},\quad t = \frac{\tau}{\lambda}.
\end{equation}
Upon considering the limit $\lambda \to 0$, we discover an AdS$_{2}$ $\times \mathbb{T}^{2}$ geometry
\begin{equation} \label{eq:metric_ext_nh}
    ds^2 = \frac{1}{3 Y^{2}} \left(-d\tau^{2} + dY^{2} \right) + 4\pi r_{0}^{2} V_{2} \left(dx_{1}^{2} + dx_{2}^{2} \right),
\end{equation}
where the radius of AdS$_{2}$ is given by $\ell_{2}^{2} = \frac{1}{3}$.


\section{Semiclassical approach to the retarded Green's function} \label{sec:classical_Green}

We begin  by reviewing the semiclassical method for computing transport coefficients in holographic setups. We work in units where $L = 16\pi G = 1$.

\subsection{Metric fluctuations and holographic shear viscosity $\left(\eta\right)$}

A direct method for computing transport coefficients in holography is based on the Kubo formula, which relates the transport coefficients to real-time correlation functions \cite{Policastro:2001yc, Policastro:2002se}. For the shear viscosity, the corresponding relation is
\begin{equation} \label{eq:Kubo_formula_eta}
    \eta = \lim_{\omega \to 0} \frac{i}{\omega} G_{x_{1}x_{2}, x_{1}x_{2}},
\end{equation}
where $G_{x_{1}x_{2},\,x_{1}x_{2}} = \langle T_{x_{1} x_{2}} T_{x_{1} x_{2}} \rangle$ is the real-time, retarded two-point function of the energy-momentum tensor operator in the dual CFT. To compute the retarded Green's function \(G_{x_{1}x_{2},\,x_{1}x_{2}}\), we consider fluctuations of the bulk metric in the \emph{shear channel},
\begin{align}
\delta g_{\mu\nu}\, dx^{\mu} dx^{\nu}
\sim
\int d\omega\, dk_1\, dk_2\;
e^{-i\omega t + i k_{1} x_1 + i k_{2} x_2}
\, g_{x_1 x_1}\, \delta g^{x_{1}}_{~x_{2}}(r)\, dx_1 dx_2 .
\end{align}
In this paper we focus on the hydrodynamic limit, in which we first take \(\vec{k} \to 0\). The linearized Einstein equation for the fluctuation  $\delta g^{x_{1}}_{~x_{2}} (r) = h(r)$ in the background geometry \eqref{metric} leads to the following radial equation with an effective  radially varying mass \(m(r)^2 = \alpha^2/r^2\) (see \cite{Hartnoll:2016tri} for further details):
\begin{equation}
\label{eq:hxy_equation_full}
\frac{1}{\sqrt{-g}} \partial_r \!\left( \sqrt{-g}\, g^{rr} \partial_r h(r) \right)
+ \left( g^{tt} \omega^2 - m(r)^2 \right) h(r) = 0 .
\end{equation}

We would like to solve this equation in the extremal background with prescribed boundary conditions corresponding to the retarded Green's function. This is achieved by solving Equation \eqref{eq:hxy_equation_full} by the method of matched asymptotic expansions. The entire spacetime geometry is divided into two regions, we closely follow the methods laid out in \cite{Edalati:2009bi, Edalati:2010hk, Edalati:2010pn} which describe the details of low-temperature computations.

In the extremal limit, the near-horizon geometry develops a long \(\mathrm{AdS}_2 \times \mathbb{T}^2\) throat. In this region the shear fluctuation behaves as a scalar field with an effective constant mass corresponding, via the AdS/CFT dictionary, to a dual operator of conformal dimension \(\Delta = 2\). This implies that at the boundary of the throat the field admits the standard normalizable and non-normalizable modes; their ratio (response/source) defines the IR Green's function.
On the other hand, since the full geometry is asymptotically AdS\(_4\), at the asymptotic boundary we obtain the usual four-dimensional Green's function, which we refer to as the UV Green's function. As a first step in this analysis we determine the relation between the UV and IR Green's functions. We will, subsequently, incorporate the quantum corrections due to coupling with the  Schwarzian modes and finally find  the quantum-corrected shear viscosity.

\subsubsection*{Solution in the near-horizon region}

In the near-horizon region, it is convenient to introduce a new coordinate \(\zeta = \frac{r - r_0}{\omega}\).
In terms of this coordinate, the ODE in Equation \eqref{eq:hxy_equation_full} takes the following  form to leading order:
\begin{equation}
h''(\zeta)
+ \left( \frac{2}{\zeta} + \ldots \right) h'(\zeta)
+ \left( \frac{1}{9 \zeta^{4}} - \frac{2}{\zeta^{2}} + \ldots \right) h(\zeta)
= 0, 
\end{equation}
where the ellipses denote higher-order terms in this expansion. The general solution is
\begin{equation} \label{near_region}
h_{\text{near}}(\zeta)
=
a_1 \left(\zeta - \frac{i}{3} \right) e^{\frac{i}{3\zeta}}
+
a_2 \left(\zeta + \frac{i}{3} \right) e^{-\frac{i}{3\zeta}} .
\end{equation}
Here the first term represents the ingoing mode, while the second term corresponds to the outgoing mode, and \(a_{1,2}\) are arbitrary constants.

\subsubsection*{Solution in the far-from-horizon region}

In the far-from-horizon region we introduce another dimensionless coordinate \(u = r/r_0\).
In this coordinate the radial equation takes the form
\begin{align}
h''(u)
+ \left( \frac{1}{u} + \frac{1}{u-1} + \frac{2}{u+2} \right) h'(u)
- \frac{\alpha^2}{r_0^2}\,
\frac{1}{u (u-1)^2 (u+2)}\, h(u)
= 0 ,
\end{align}
where we have neglected the term proportional to \(\omega^2\).
For a massless scalar field the third term vanishes and the solution simplifies (see e.g.~\cite{PandoZayas:2025snm}).
For a generic massive probe field the equation is of Heun type and the solutions are given in terms of Heun functions.
However, in our case at extremality we have \(\alpha = \sqrt{6}\, r_0\), and the solution simplifies to
\begin{equation}\label{far_region}
h_{\text{far}}(u)
=
\frac{1}{8 (u-1)^2}
\Big[
\big(u^2 - 2u - 2\big)
\left(8c_1 + c_2 \log\!\frac{u}{u+2}\right)
+ 2c_2 (u-3)
\Big] .
\end{equation}
%
\subsection*{Matching and retarded Green's function $G_{xy,xy}^{R}$}

So far we have obtained solutions to the linearized equations of motion in the two asymptotic regions of the spacetime. To fix the undetermined constants \(c_1\) and \(c_2\), we match the solutions in the overlap region. For this purpose, we expand the near-horizon solution for large values of \(\zeta\) and the far-from-horizon region solution for small values of  \(u\), and then match the two expansions to determine the relations between \(a_1, a_2\) and \(c_1, c_2\).\\

\noindent Expanding the near-horizon solution \eqref{near_region} in the overlap region, we obtain
\begin{equation}
h_{\text{near}} (\zeta)
=
\zeta \, (a_1+a_2)
\left( 1+O \left(\frac{1}{\zeta^2} \right)\right)
+
\frac{i}{81 \zeta^2} (a_1-a_2)
\left( 1+O \left(\frac{1}{\zeta^2} \right)\right) .
\end{equation}
Imposing the ingoing boundary condition at the horizon sets \(a_2=0\), and therefore
\begin{align}\label{near_overlap}
h_{\text{near}}
=
a_1 \left( (r-r_0)
+ \mathcal{G}_{\text{IR}} (\omega) \frac{1}{(r-r_0)^2}
+ \ldots \right) .
\end{align}
The above expression coincides with the behavior of a neutral scalar field of conformal dimension \(\Delta=2\) near the boundary of an \(\mathrm{AdS}_2\) geometry. This is consistent with the fact that the effective mass squared of the scalar field near the horizon is, \(m^{2} = \alpha^{2}/r_0^{2} = 6\), which corresponds to \(m^2 L_2^2 = 2\).
For later convenience, we define the IR retarded Green's function as the ratio of the coefficients of the normalizable and non-normalizable modes
\begin{equation}\label{IR_green}
\mathcal{G}_{\text{IR}} (\omega)
=
\frac{i \omega^3}{81} .
\end{equation}
%
The behavior of the far-from-horizon solution \eqref{far_region} in the overlap region is
\begin{align}\label{far_overlap}
h_{\text{far}}
&=
-3 r_0^2
\left(c_1+ \frac{c_2}{24} (4-3 \log 3) \right)
\frac{1}{(r-r_0)^2}
\Big(1+O((r-r_0)^2)\Big)
\nonumber \\
&\quad
-\frac{1}{27r_0} c_2 (r-r_0)
\Big(1+O(r-r_0)\Big) .
\end{align}
Matching \eqref{near_overlap} with \eqref{far_overlap} gives
\begin{align}
c_2 =-27r_0\, a_1 ,\quad
c_1 = a_1 \left( \frac{9}{8} r_0 (4-3 \log 3)
-\frac{\mathcal{G}_{\text{IR}} \left(\omega \right)}{3r_0^2} \right) .
\end{align}

\subsubsection*{Relation between UV and IR Green's functions}

To obtain the relation between the UV and IR Green's functions, we first
expand Equation \eqref{far_region} near the AdS$_{4}$ boundary. This yields
\begin{equation}
\lim_{u \rightarrow \infty} h_{\text{far}}
\sim
c_1
-
\left(6 c_1 + \frac{c_2}{3} \right)
\frac{r_0^3}{r^3}
+ \ldots .
\end{equation}
which corresponds to the behavior of a massless scalar field in AdS$_{4}$, and is consistent with the fact that near the boundary the effective mass satisfies $\alpha/r \rightarrow 0$. The first term corresponds to the non--normalizable mode, while the second term corresponds to the normalizable mode. Therefore, the four-dimensional UV retarded Green's function is
\begin{equation}
G_{\text{UV}}
=-\mathcal N_\eta\left[
-
\frac{r_0^3}{c_1}
\left(6 c_1+\frac{c_2}{3} \right)\right]
=-\mathcal N_\eta\left[-
\frac{c_2}{3c_1} r_0^3\right],
\end{equation}
where we neglect the $c_i$-independent constant part. 
The on-shell action normalization gives a prefactor of 
\begin{equation}
    \mathcal N_\eta=\frac{3}{16\pi GL^4}.
\end{equation}
The derivation of this normalization can be found in Appendix \ref{app:normalization-boundary-coefficients}. In the following we set $L=1$ and $16 \pi G = 1$ for simplicity. 
Using Equation \eqref{IR_green}, this can be expressed as
%
%
\begin{equation} \label{eq:UV_IR_eta}
 G_{\text{UV}} =  \frac{648 \, r_0^6}
 {8\, \mathcal{G}_{\text{IR}} \left(\omega \right)-27 r_0^3 (4-3 \log 3)}.
 \end{equation}
Substituting the expression for $\mathcal{G}_{\text{IR}}$ from Equation \eqref{IR_green}, we obtain
%
\begin{align}
G_{\text{UV}}
=-
\frac{24 r_0^3}{4-3 \log 3}
-
\frac{64 \,  i \omega ^3}{729 (4-3 \log 3)^2}
+
\mathcal{O}\!\left(\omega ^6\right).
\end{align}
This implies
%

\begin{align}
\text{Im}\, G_{\text{UV}}
=-
\frac{64 \, \omega ^3}{729 (4-3 \log 3)^2}
+
\mathcal{O}\!\left(\omega ^6\right),
\end{align}
The shear viscosity can be obtained from the UV Green's function using the Kubo formula \eqref{eq:Kubo_formula_eta}, which leads to
%
\begin{equation}
\eta =- \lim_{\omega \rightarrow 0} \frac{1}{\omega} \operatorname{Im} G_{\mathrm{UV}} = \lim_{\omega \rightarrow 0}
\frac{ 64 \times 3r_0^2 \omega ^2}{729 (4-3 \log 3)^2}
\rightarrow 0 .
\end{equation}
%
Therefore, the shear viscosity $\eta$ vanishes at extremality. To obtain more detailed information about how $\eta$ approaches zero as $T \rightarrow 0$, one could, in principle, solve the radial equation in the near-extremal
geometry rather than the extremal one. This leads to a Heun-type equation in the near-horizon region, which is technically cumbersome, although it
can in principle be treated using the methods developed in
\cite{Bonelli:2022ten, Das:2024fwg}. A simpler approach is to follow the procedure of \cite{Hartnoll:2016tri}. Rather than repeating the analysis here, we quote their result
\begin{equation}
\frac{4\pi \eta}{s}
=
\frac{512\pi^{2}}{2187 \left(\frac{4}{3}-\log 3\right)^{2}}
\left(\frac{T}{\alpha}\right)^{2}
+
\mathcal{O}\!\left(\frac{T}{\alpha}\right)^{4}.
\end{equation}
The shear viscosity to entropy density ratio vanishes quadratically in temperature. It fits the entropy production interpretation advanced in  \cite{Hartnoll:2016tri} as a result of far IR breaking of translational symmetry.

\subsection{Gauge field fluctuations and holographic DC conductivity $\left(\sigma_{\text{DC}}\right)$}

We now consider the dynamics of a U$(1)$ gauge field in the four-dimensional asymptotically AdS$_4$ geometry \eqref{metric} in the probe approximation. The action for the gauge field is the standard Maxwell action
\begin{equation}
    I_{\mathrm{Maxwell}} = -\frac{1}{4}\int d^{4}x \sqrt{-g}\, F^{\mu\nu} F_{\mu\nu},\quad F_{\mu\nu} = \nabla_{\mu}A_{\nu} - \nabla_{\nu} A_{\mu}.
\end{equation}
In order to compute the holographic DC conductivity, we need the two-point correlation function $\langle J_{\mu} J_{\nu} \rangle$ of a conserved current $J_{\mu}$ in the dual field theory. The gauge/gravity duality relates the conserved current $J_{\mu}$ in the boundary to the fluctuation of the bulk gauge field $A_{\mu}$. We consider the following kind of fluctuation
\begin{equation}
    A_{\mu} = \lbrace 0, 0, e^{-i \omega t} a_{x_{1}}(r), e^{-i \omega t} a_{x_{2}}(r) \rbrace,
\end{equation}
where, once again, we have considered $\vec{k} = 0$ from the very beginning. The equation of motion for $a_{x_{1}}$ is given by
\begin{equation}
    a_{x_{1}}''(r) + \frac{f'(r)}{f(r)} a_{x_{1}}'(r) + \frac{\omega^2}{ f(r)^{2}} a_{x_{1}}(r) = 0.
\end{equation}
The other spatial component $a_{x_{2}}$ obeys an identical equation. For the physical questions we address in this paper, it is sufficient to work with only one component, and, accordingly, we set $a_{x_{2}} = 0$.

We will now solve the equation of motion and compute the retarded Green's function $G_{x_{1}x_{1}}$ by the method of matched asymptotics.

\subsubsection*{Solution in the near-horizon region}
In the region of spacetime close to the outer horizon $r_{0}$, we use the coordinate $\zeta = \frac{r - r_{0}}{\omega}$. Since we are interested in a small frequency limit, let us expand the fluctuation field in a power series in $\omega$:
\begin{equation} \label{eq:small_freq_a}
    a_{x_{1}} \left(\zeta\right) = a_{x_{1}}^{(0)} \left(\zeta\right) + \omega\,a_{x_{1}}^{(1)} \left(\zeta\right) + \mathcal{O}\left(\omega^{2}\right).
\end{equation}
Then the leading order linearized equation of motion takes on a very simple form
\begin{equation} \label{eq:ax_near_eom}
    \frac{d^{2}a_{x_{1}}^{(0)}}{d\zeta^{2}} +\frac{2}{\zeta } \frac{da_{x_{1}}^{(0)}}{d\zeta} +\frac{a_{x_{1}}^{(0)} \left(\zeta\right)}{9 \zeta ^4} = 0.
\end{equation}
The general solution to this equation is given by
\begin{equation}
    a_{x_{1},\,\mathrm{near}}^{(0)}\left(\zeta\right) = a_{1} e^{\frac{i}{3\zeta}} + a_{2} e^{-\frac{i}{3\zeta}}.
\end{equation}
As usual, the second term here corresponds to the outgoing mode. Hence we put $a_{2} = 0$ upon imposing the ingoing boundary condition at the horizon. Near the overlapping region $\left(\zeta \to \infty\right)$ we may expand the solution as
\begin{equation} \label{eq:ax_nearsol_overlap}
    \left.a_{x_{1},\,\mathrm{near}}^{(0)}\right|_{\zeta\to\infty} = a_{1} + \frac{i a_{1}}{3\zeta} - \frac{a_{1}}{18 \zeta^{2}}.
\end{equation}

Let us also note that the Equation \eqref{eq:ax_near_eom} is exactly the Klein-Gordon equation of a neutral, massless scalar field in the near-horizon AdS$_{2}$ geometry. Therefore, in this region, we may assume that the dynamics of the gauge field fluctuation $a_{x_{1}}$ is identical to a primary scalar operator in the CFT$_{1}$ dual to the AdS$_{2}$, with conformal dimension $\Delta = 1$. We will use this fact in the next section to work out the quantum-corrected retarded two-point correlator.

\subsubsection*{Solution in the far-from-horizon region}

We use the dimensionless coordinate $u = r/r_{0}$ in the asymptotically AdS$_{4}$ region of the spacetime as before. Employing a power series expansion in small frequency  $\omega$ similar to Equation \eqref{eq:small_freq_a}, we obtain the leading order ODE
\begin{equation}
    \frac{d^{2}a_{x_{1}}^{(0)}}{du^{2}} + \frac{2 \left(u^2+u+1\right)}{(u-1) u (u+2)} \frac{d a_{x_{1}}^{(0)}}{du} = 0,
\end{equation}
with the general solution
\begin{equation}
    a_{x_{1},\,\mathrm{far}}^{(0)}(u) = c_{1} + c_{2} \left(-\frac{1}{3\left(u-1\right)} - \frac{4}{9} \tanh^{-1}{\frac{2u+1}{3}} \right).
\end{equation}
Near the overlapping region $\left(u \to 1\right)$, the solution admits the expansion
\begin{equation} \label{eq:ax_farsol_overlap}
    \left.a_{x_{1}}^{(0)}(u) \right|_{u \to 1} = -\frac{c_{2}}{3 \left(u-1\right)} + \left(c_{1} + \frac{2 \pi i c_{2}}{9} + \frac{2 c_{2}}{9} \log \left(\frac{u-1}{3} \right) + \cdots \right),
\end{equation}
where ellipses denote terms that vanish as $u \to 1$. On the other hand, the asymptotic behavior of the far-region solution near the AdS$_{4}$ conformal boundary is seen to be given by
\begin{equation}\label{eq:far1}
    \left. a_{x_{1}}^{(0)} (u) \right|_{u \to \infty} = c_{1} + \frac{2\pi i\,c_{2}}{9} - \frac{c_{2}}{u},
\end{equation}
which fulfills the expectation for a dual conserved current operator $\Delta_{3d}=2$.
\subsubsection*{Matching and the retarded Green's function $G^{R}_{x_{1},x_{1}}$}

Let us rewrite the asymptotic forms of the near and far region solutions in eqs.~\eqref{eq:ax_nearsol_overlap} and \eqref{eq:ax_farsol_overlap} in terms of the original radial coordinate $r$:
\begin{subequations}
\begin{align}
\left. a_{x_{1},\,\mathrm{near}}^{(0)} (r) \right|_{r \to \infty}
&=
a_{1}
+
\frac{i\omega a_{1}}{3\left(r - r_{0}\right)}
+
\mathcal{O}\!\left(\frac{1}{(r-r_{0})^{2}}\right),
\label{eq:ax_matching_1}
\\
\left. a_{x_{1},\,\mathrm{far}}^{(0)} (r) \right|_{r \to r_{0}}
&=
- \frac{c_{2} r_{0}}{3\left(r - r_{0}\right)}
+
c_{1}
+
\frac{2\pi i\, c_{2}}{9}
+
\frac{2 c_{2}}{9}
\log \!\left(\frac{r - r_{0}}{3 r_{0}} \right)
+
\mathcal{O}\!\left(r - r_{0}\right).
\label{eq:ax_matching_2}
\end{align}
\end{subequations}
In the extremal limit the geometry develops an AdS$_{2}\times \mathbb{T}^{2}$ throat. We have already observed that the gauge field fluctuation behaves as a massless scalar field in the two–dimensional AdS space, corresponding to $\Delta=1$. As a result, one can define an IR Green's function $\mathcal{G}_{\text{IR}}$, as before, as the ratio of the normalizable and non–normalizable modes. With this definition,
Equation \eqref{eq:ax_matching_1} can be written as
\begin{equation}
\label{eq:ax_matching_11}
\left. a_{x_{1},\,\mathrm{near}}^{(0)} (r) \right|_{r \to \infty}
=
a_{1}
\left(
1 +
\frac{\mathcal{G}_{IR}\left(\omega \right)}{(r - r_{0})}
+ \ldots
\right),
\qquad
\mathcal{G}_{\text{IR}} \left(\omega \right)=\frac{i\omega}{3}.
\end{equation}
Matching eqs.~\eqref{eq:ax_matching_11} and \eqref{eq:ax_matching_2} we obtain
\begin{equation}
c_{2}
=
- a_{1}\frac{3\,\mathcal{G}_{\mathrm{IR}} \left(\omega \right)}{r_{0}},
\qquad
c_{1}
=
a_{1}\!\left(
1+\frac{2 i \pi}{3 r_{0}}\mathcal{G}_{\mathrm{IR}} \left(\omega \right)
\right).
\end{equation}
From the near-boundary behavior of the fluctuations in Equation \eqref{eq:far1}, the UV Green's function can be defined and written in terms of IR Green's function as
\begin{equation}\label{matching_green_sigma}
G_{\text{UV}}
={\textcolor{red}{-}}\mathcal N_\sigma\left[
-\frac{c_{2} r_{0}}
{c_{1}+\frac{2 i \pi c_{2}}{9}}\right]
=
{\color{red}-}3\,\mathcal{G}_{IR} \left(\omega \right),
\end{equation}
where the normalization coefficient $\mathcal N_\sigma$ is calculated in Appendix \ref{app:normalization-boundary-coefficients}. With $\tfrac{1}{16\pi G}=L=g_F=1$, it is just 1.
The conductivity can be obtained from the Kubo formula 
\begin{equation}
    \sigma_{\text{DC}} = \lim_{\omega \to 0} \frac{i}{\omega} G_{x_{1}x_{1}}.
\end{equation}
Upon substituting $G_{x_{1}x_{1}} = G_{\text{UV}}$, we arrive at
\begin{equation}
\label{eq:sigam_zero_gauge_field}
    \sigma_{\text{DC}} = 1.
\end{equation}
The DC conductivity is, therefore, finite. This result is expected from the physics of momentum relaxation  \cite{Andrade:2013gsa}. In fact, the result coincides with the computation in massive gravity \cite{Blake:2013bqa}, provided the momentum relaxation parameter $\alpha$ is identified with the graviton mass. Moreover, the $\left(3+1\right)$-dimensional bulk result for $\sigma_{\text{DC}}$ is temperature-independent, although temperature dependence enters in higher dimensions \cite{Andrade:2013gsa}.

\section{Quantum-corrected Green's function} \label{sec:quantum_Green}

\begin{figure}[htb!]
    \centering
    \includegraphics[scale = 1]{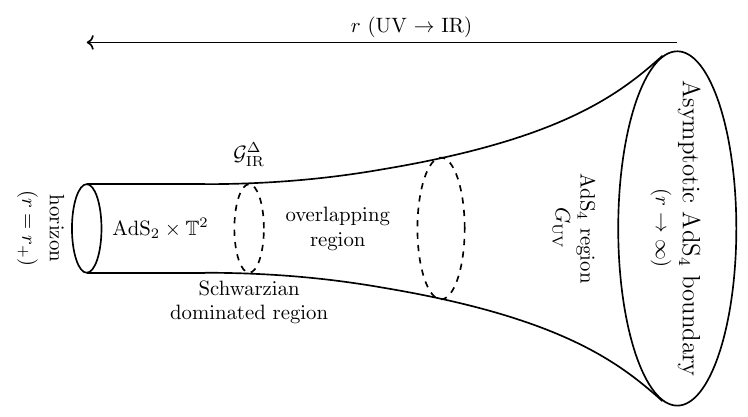}
    \caption{A schematic view of the near-extremal black brane geometry. A long AdS$_{2} \times \mathbb{T}^{2}$ throat appears close to the outer horizon $\left(r \to r_{+} \right)$. Far from the horizon $\left(r \to \infty \right)$, the spacetime asymptotes to a four-dimensional anti-de Sitter geometry. The two regions are smoothly connected through an `overlapping region' where the matched asymptotic expansion is performed in the semiclassical analysis. The Schwarzian modes get strongly coupled at sufficiently low temperature, and dominate the near-horizon physics.}
    \label{fig:schematic_geometry}
\end{figure}

In this section we incorporate the quantum fluctuations that are localized in the long throat of the spacetimes at very low temperatures. As illustrated in Figure \ref{fig:schematic_geometry}, we are going to couple the operator ${\cal O}_\Delta$ to the Schwarzian theory in a version of nAdS$_2$/nCFT$_1$. The conformal dimension in the Schwarzian nCFT$_1$ is determined by the mass of the bulk perturbation. 

For the application of Kubo formulas, we are ultimately interested in computing the imaginary part of the retarded Green's function in momentum space. We begin, however, with the exact Wightman function and show how it is related to the retarded Green's function.

The position-space two-point Wightman function \(\left\langle O_{\Delta}(t)\, O_{\Delta}(0) \right\rangle\) of the Schwarzian theory is given by \cite{Mertens:2022irh}
\begin{equation}
\begin{aligned}
\langle O_{\Delta}(t) O_{\Delta}(0) \rangle
&=
\frac{e^{S_0}}{Z(T)}
\frac{1}{8\pi^{4}(2C)^{2\Delta}\Gamma(2\Delta)}
\\
&\quad \times
\int_{0}^{\infty}
\prod_{i=1}^{2}
\left(
dk_i^{2}\,\sinh(2\pi k_i)
\right)
e^{- i t \frac{k_1^{2}}{2C}-(\beta-it)\frac{k_2^{2}}{2C}}
\prod_{\sigma_1,\sigma_2=\pm 1}
\Gamma\!\left(\Delta+i\sigma_1 k_1+i\sigma_2 k_2\right).
\end{aligned}
\end{equation}
The frequency-space Wightman function is then\footnote{
Here \(G_{\Delta}(\omega)\) is the momentum-space Fourier transform of the Wightman function
\begin{equation*}
G_{\Delta}(\omega)
= \frac{1}{\pi}
\int_{-\infty}^{\infty} dt \,
e^{i\omega t}
\left\langle O_{\Delta}(t)\, O_{\Delta}(0) \right\rangle .
\end{equation*}
To obtain \eqref{sw_correlator_k}, we have also used the Fourier transform
\begin{equation*}
\frac{1}{\pi}
\int_{-\infty}^{\infty} dt \,
e^{i\omega t}
e^{- i t \frac{k_1^{2}-k_2^{2}}{2C}}
=
2\,\delta\!\left(\omega - \frac{k_1^{2}-k_2^{2}}{2C}\right)
=
4C\,\delta\!\left(k_1^{2} - (k_2^{2}+2C\omega)\right).
\end{equation*}}
\begin{equation}\label{sw_correlator_k}
\begin{aligned}
G_{\Delta}(\omega)
&=
\frac{e^{S_0}(2C)}{Z(T)\,4\pi^{4}(2C)^{2\Delta}\Gamma(2\Delta)}
\int_{0}^{\infty} dk \;
2k \,
e^{-\frac{k^{2}}{2CT}}
\sinh(2\pi k)
\\
&\quad \times
\sinh\!\left(2\pi\sqrt{k^{2}+2C\omega}\right)
\prod_{\sigma_1,\sigma_2=\pm1}
\Gamma\!\left(
\Delta
+i\sigma_1\sqrt{k^{2}+2C\omega}
+i\sigma_2 k
\right).
\end{aligned}
\end{equation}
%

\subsection{Stress tensor correlator and shear viscosity $\eta$}

For \(\Delta=2\), the product of gamma functions simplifies and we obtain
\begin{equation}
G_{\Delta=2}(\omega)=G_1(\omega)+G_2(\omega) .
\end{equation}
where
\begin{equation}
\begin{aligned}
G_1(\omega)
&=
\frac{e^{S_0}\,\omega(1+2 C \omega)^2}{Z(T)\,48 C^2\pi^{2}}
\int_{0}^{\infty} dk \;
k\, e^{-\frac{k^{2}}{2CT}}
\sinh(2\pi k) \\
&\quad \times
\frac{
\sinh\!\left(2\pi\sqrt{k^{2}+2C\omega}\right)
}{
\sinh\!\left[\pi\!\left(\sqrt{k^{2}+2C\omega}+k\right)\right]
\sinh\!\left[\pi\!\left(\sqrt{k^{2}+2C\omega}-k\right)\right]
},
\end{aligned}
\end{equation}
and
\begin{equation}
\begin{aligned}
G_2(\omega)
&=
\frac{e^{S_0}\,\omega}{Z(T)\,12 C^2\pi^{2}}
\int_{0}^{\infty} dk \;
k^3\, e^{-\frac{k^{2}}{2CT}}
\sinh(2\pi k) \\
&\quad \times
\frac{
\sinh\!\left(2\pi\sqrt{k^{2}+2C\omega}\right)
}{
\sinh\!\left[\pi\!\left(\sqrt{k^{2}+2C\omega}+k\right)\right]
\sinh\!\left[\pi\!\left(\sqrt{k^{2}+2C\omega}-k\right)\right]
}.
\end{aligned}
\end{equation}
To evaluate $G_{1,2} (\omega)$ we closely follow \cite{Kanargias:2025vul}. Let us first rewrite $G_1(\omega)$ as
\begin{equation}\label{green_int1}
G_1(\omega)
=
\frac{e^{S_0}\,\omega (1+2 C \omega)^2}{Z(T)\,48 C^2 \pi^{2}}
\int_{0}^{\infty} dk \;
g_T(k)\, f(k,\omega),
\end{equation}
where
\begin{align}
g_T(k) &= k\, e^{-\frac{k^{2}}{2CT}} \sinh(2\pi k),
\\
f(k,\omega) &=
\frac{
\sinh\!\left(2\pi\sqrt{k^{2}+2C\omega}\right)
}{
\sinh\!\left[\pi\!\left(\sqrt{k^{2}+2C\omega}+k\right)\right]
\sinh\!\left[\pi\!\left(\sqrt{k^{2}+2C\omega}-k\right)\right]
}. \label{eq_fk}
\end{align}
We work in the regime in which \(\omega\) is the smallest scale in the system, so that \(C\omega \ll CT\) and similarly for the other scales. In this limit \(f(k,\omega)\) can be approximated as (see the appendix of \cite{Kanargias:2025vul} for details)
\begin{equation}\label{fk_approx}
f(k, \omega) \sim \frac{k}{\pi C \omega}+ \ldots 
\end{equation}
This implies that the integration in \eqref{green_int1} can be approximated as
\begin{align}\label{int_I1}
I_1(\omega, T)
&=\int_{0}^{\infty} dk \; g_T(k)\, f(k,\omega)\nonumber \\
&=\int_{0}^{\infty} dk \; \frac{k^2}{ \pi C \omega} \; e^{-\frac{k^{2}}{2CT}} \sinh(2\pi k) \nonumber \\
& =\frac{(2\pi C)^{3/2}}{\omega}\, T^{5/2}\, e^{2\pi^{2}CT}
\left[
\operatorname{erf}\!\left(\sqrt{2\pi^{2}CT}\right)
\left(1+\frac{1}{4\pi^{2}CT}\right)
+ \frac{e^{-2\pi^{2}CT}}{\sqrt{2\pi^{3}CT}}
\right].
\end{align}
Similarly, for \(G_2(\omega)\) we obtain
\begin{align}\label{int_I2}
I_2(\omega, T)
&= \int_{0}^{\infty} dk \;
\frac{k^{4}}{\pi C \omega}\,
e^{-\frac{k^{2}}{2CT}}
\sinh(2\pi k)
\nonumber\\
&= \frac{(2\pi C)^{7/2}}{\omega}\,
T^{9/2}\,
e^{2\pi^{2}CT}
\Bigg[
\operatorname{erf}\!\left(\sqrt{2\pi^{2}CT}\right)
\left(
1+\frac{3}{2\pi^{2}CT}
+\frac{3}{16\pi^{4}(CT)^{2}}
\right)
\nonumber\\
&\hspace{3.2cm}
+ \frac{e^{-2\pi^{2}CT}}{\sqrt{2\pi^{3}CT}}
\left(
1+\frac{5}{4\pi^{2}CT}
\right)
\Bigg] .
\end{align}
Using the fact that the partition function is
\begin{equation}\label{eq:partition_function1}
Z(T)=\frac{(C T)^{3/2}}{\sqrt{2 \pi}} e^{S_0+2\pi^2 C T},
\end{equation}
we obtain
\begin{equation}
G_1(\omega)
=
\frac{(1+2 C \omega)^2 T}{12 C^2}
\left[
\operatorname{erf}\!\left(\sqrt{2\pi^{2}CT}\right)
\left(1+\frac{1}{4\pi^{2}CT}\right)
+ \frac{e^{-2\pi^{2}CT}}{\sqrt{2\pi^{3}CT}}
\right],
\end{equation}
and
\begin{align}
G_2(\omega)
& =
\frac{4\pi^2}{3}T^3
\Bigg[
\operatorname{erf}\!\left(\sqrt{2\pi^{2}CT}\right)
\left(
1+\frac{3}{2\pi^{2}CT}
+\frac{3}{16\pi^{4}(CT)^{2}}
\right)
\nonumber\\
&\hspace{4.8cm}
+ \frac{e^{-2\pi^{2}CT}}{\sqrt{2\pi^{3}CT}}
\left(
1+\frac{5}{4\pi^{2}CT}
\right)
\Bigg] .
\end{align}
The IR Wightman correlator in the hydrodynamic limit $(\omega \rightarrow 0)$ takes the form
\begin{align}\label{eq:IR_Wight_Final}
G_{\Delta=2}
&=
\frac{CT}{12C^3}
\Bigg[
\operatorname{erf}\!\left(\sqrt{2\pi^{2}CT}\right)
\left(
1+\frac{1}{4\pi^{2}CT}
\right)
+
\frac{e^{-2\pi^{2}CT}}{\sqrt{2\pi^{3}CT}}
\nonumber\\
&\hspace{1cm}
+
16\pi^2(CT)^2
\Bigg(
\operatorname{erf}\!\left(\sqrt{2\pi^{2}CT}\right)
\left(
1+\frac{3}{2\pi^{2}CT}
+\frac{3}{16\pi^{4}(CT)^{2}}
\right)
\nonumber\\
&\hspace{4.2cm}
+
\frac{e^{-2\pi^{2}CT}}{\sqrt{2\pi^{3}CT}}
\left(
1+\frac{5}{4\pi^{2}CT}
\right)
\Bigg)
\Bigg].
\end{align}
Writing Eq.~\eqref{eq:UV_IR_eta} we have used the retarded IR correlator, which corresponds to imposing ingoing boundary conditions at the horizon, whereas the Schwarzian correlator $G_{\Delta}$ is a Wightman correlator. The imaginary part of the retarded correlator is related to the Wightman correlator through
\begin{equation}\label{eq:reln_Wight_Retarded}
\operatorname{Im} G_{\Delta}^{R}(\omega)
=
\frac{1}{2}
\left(
1-e^{-\beta\omega}
\right)
G_{\Delta}(\omega)
\approx
\frac{\beta\omega}{2}\,
G_{\Delta}(\omega),
\end{equation}
where we have assumed the regime of $\omega\to 0$ at fixed, albeit small, temperature. 

The real part of the retarded correlator is then determined from the Kramers--Kronig relation,
\begin{equation}
\mathrm{Re}\,G^R_{\Delta}(\omega)
=
\frac{1}{\pi}\,
\mathcal{P}
\!\int_{-\infty}^{\infty}
d\omega'\,
\frac{
\mathrm{Im}\,G^R_{\Delta}(\omega')
}{
\omega'-\omega
},
\end{equation}
where $\mathcal{P}$ denotes the principal value. It follows from eqs.~\eqref{eq:IR_Wight_Final} and \eqref{eq:reln_Wight_Retarded} that the imaginary part of the retarded Green's function is linear in $\omega$. Consequently, the real part vanishes up to an additive constant corresponding to a contact term, which does not carry any physical information. Therefore, the retarded correlator takes the form
\begin{equation}\label{eq:retarded_green}
G_{\Delta}^{R}
= i \operatorname{Im} G_{\Delta}^{R}=
i\,
\frac{\beta\omega}{2}\,
G_{\Delta}.
\end{equation}
Substituting this expression in place of $\mathcal{G}_{\mathrm{IR}}$ in eq.~\eqref{eq:UV_IR_eta}, and subsequently extracting the imaginary part, we obtain the following expression for the shear viscosity
\begin{align}\label{eq:eta_final_linearaxion}
\eta
&=
-\lim_{\omega\rightarrow0}
\frac{1}{\omega} \operatorname{Im}
G^{R}_{\mathrm{IR}}
\nonumber\\
&=
\frac{2 \, 
e^{-2\pi^{2}CT}
}{
27 \pi^{2}C^{2}CT(4-3\log 3)^{2}
}
\Bigg[
e^{2\pi^{2}CT}
\Big(1+
2 (9+2\pi^2) CT+ 144\pi^2 (CT)^2+
96\pi^{4}(CT)^{3}
\Big)
\nonumber\\
&\hspace{4cm}
\times
\operatorname{erf}\!\left(
\sqrt{2}\pi\sqrt{CT}
\right)
+
2\sqrt{2\pi CT}
\Big(1+ 30 CT+24\pi^2 (CT)^2
\Big)
\Bigg].
\end{align}
For convenience, we present the asymptotic behavior of $\eta$ in the limits of small and large $CT$:
\begin{align}
\eta
=
\begin{cases}
\displaystyle
\frac{8 \sqrt{2}
}{27 \pi^{3/2}C^{2}(4-3\log 3)^{2}
}
\,
\frac{1}{\sqrt{CT}}
\left[
1+\frac{2}{3}(36+\pi^{2})CT+\cdots
\right],
& \qquad CT \ll 1,
\\[4mm]
\displaystyle
\frac{64 \pi^{2}
}{9 \, C^2 \, 
(4-3\log 3)^{2}
}
\,
(CT)^{2}
\left[
1+\frac{3}{2\pi^{2}CT}+\cdots
\right],
& \qquad CT \gg 1.
\end{cases}
\end{align}





\begin{figure}[t]
    \centering
    \includegraphics[width=0.8\linewidth]{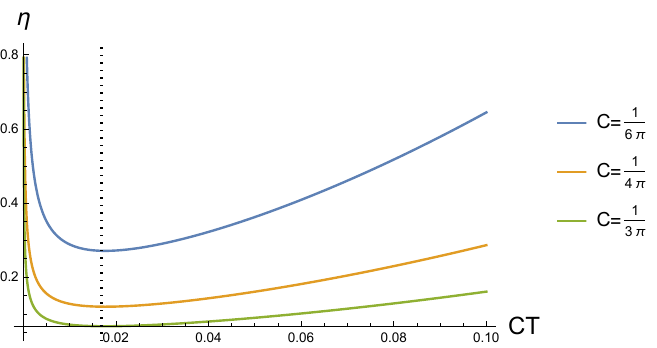}
    \caption{Behavior of $\eta$, given by Equation \eqref{eq:eta_final_linearaxion}, as a function of $CT$ for various values of the Schwarzian coupling parameter $C$. The transverse volume is set to $V_2=1$. The vertical line indicates the position of the minimum, located at $CT_{\mathrm{min}} \approx 0.01769$, which is independent of $C$.}
    \label{eta_linear_axion}
\end{figure}

In Figure~\ref{eta_linear_axion}, we present the behavior of $\eta$ as a function of $CT$ for various values of the Schwarzian coupling $C$. It is worth highlighting that, for $\Delta=1$, all the $C$-dependence of the correlator (and hence of $\eta$) enters solely through the dimensionless combination $CT$. In contrast, for $\Delta \neq 1$, the correlator depends explicitly on $C$ in addition to $CT$. The corresponding power of $C$ is completely determined by the conformal dimension $\Delta$.
The figure shows that, unlike the $\Delta=1$ case, $\eta$ diverges in both the low- and high-$CT$ regimes.
In particular, the leading divergence in the large-$CT$ regime is quadratic, whereas in the $CT \ll 1$ regime the divergence behaves as $1/\sqrt{CT}$, similar to the $\Delta=1$ case. Moreover, $\eta$ exhibits a local minimum located at
\begin{equation}
    CT_{\mathrm{min}} \approx 0.01769.
\end{equation}

\begin{figure}[t]
    \centering
    \includegraphics[width=0.7\linewidth]{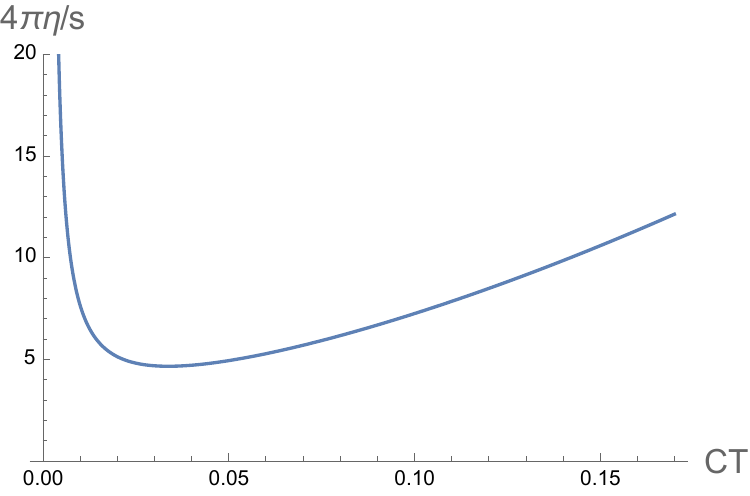}
    \caption{Quantum-corrected $\eta/s$ as a function of $CT$ for $r_0=1, V_2=3/4$.}
        \label{eta_s_linear_axion}
\end{figure}

To study the behavior of $\eta/s$, one must take into account the quantum-corrected entropy density arising from the Schwarzian-corrected partition function. This correction was derived in \cite{Iliesiu:2020qvm} (see also \cite{Iliesiu:2022onk, Banerjee:2023gll, Kapec:2023ruw, Rakic:2023vhv, Maulik:2024dwq}) and has been shown to be fairly universal across various asymptotics and dimensions in \cite{PandoZayas:2026vbg}. The corrected entropy density is given by
\begin{equation}\label{eq:corrected_s}
s=s_0+\frac{3}{2V_2}\log CT+O(T),
\end{equation}
where $s_0$ denotes the extremal entropy density, given in Equation \eqref{eq:some_relations}. This relation indicates that one cannot take the $CT \to 0$ limit arbitrarily far, since for sufficiently small $CT$ the entropy becomes negative, which is clearly unphysical. This signals the breakdown of the Schwarzian correction, occurring at
\begin{equation}
(CT)_{*}\sim \exp\left(-\frac{2S_0}{3}\right).
\end{equation}
Below this scale, the discreteness of the black hole microstate spectrum becomes important, and the effective Schwarzian description is no longer reliable. In Figure~\ref{eta_s_linear_axion}, we present the behavior of $\eta/s$ as a function of $CT$.
We note that the quantum-corrected form of $\eta/s$ in the presence of momentum relaxation differs from the results obtained previously \cite{PandoZayas:2025snm,Cremonini:2025yqe,Gouteraux:2025exs,Kanargias:2025vul} where the value of $1/(4\pi)$ still figured prominently. In particular, the behavior obtained in this manuscript looks more like realistic materials (see a relevant discussion in  \cite{Cremonini:2012ny}). We plan to explore this similitude in more detail elsewhere.

\subsection{Current-current correlator and $\sigma_{\mathrm{DC}}$}

As discussed previously, the fluctuation relevant for computing the
current--current correlator behaves as a massless scalar field in the
IR AdS$_{2}\times \mathbb{T}^{2}$ throat. Consequently, we are interested in the Schwarzian-corrected momentum-space correlator \eqref{sw_correlator_k} for $\Delta=1$, which takes
a particularly simple form  recently discussed in \cite{Kanargias:2025vul}
\begin{equation}\label{green1_sigma}
\begin{aligned}
G_{\Delta=1}(\omega)
&=
\frac{e^{S_0}\omega}{Z(T)\, 2 \pi^{2}}
\int_{0}^{\infty} dk \;
k\, e^{-\frac{k^{2}}{2CT}}
\sinh(2\pi k) \\
&\quad \times
\frac{
\sinh\!\left(2\pi\sqrt{k^{2}+2C\omega}\right)
}{
\sinh\!\left[\pi\!\left(\sqrt{k^{2}+2C\omega}+k\right)\right]
\sinh\!\left[\pi\!\left(\sqrt{k^{2}+2C\omega}-k\right)\right]
}.
\end{aligned}
\end{equation}
As before, we are interested in the regime \(C\omega \ll CT\), where Equation \eqref{green1_sigma} reduces to
\begin{align}
G_{\Delta=1}(\omega)
=
\frac{e^{S_0}\omega}{Z(T)\, 2 \pi^{2}}
I_1(\omega,T),
\end{align}
where \(I_1(\omega,T)\) is defined in Equation \eqref{int_I1}. This yields the Schwarzian-corrected IR Wightman Green's function
\begin{equation}\label{eqn:Green_delta1}
G_{\Delta=1}(\omega)
=
2 T
\left[
\operatorname{erf}\!\left(\sqrt{2\pi^{2}CT}\right)
\left(1+\frac{1}{4\pi^{2}CT}\right)
+
\frac{e^{-2\pi^{2}CT}}{\sqrt{2\pi^{3}CT}}
\right].
\end{equation}
As before, the imaginary part of the retarded Green's function can be extracted using \eqref{eq:reln_Wight_Retarded}, which gives
\begin{equation}
   \operatorname{Im} G^R_{\Delta=1}(\omega)=\omega
\left[
\operatorname{erf}\!\left(\sqrt{2\pi^{2}CT}\right)
\left(1+\frac{1}{4\pi^{2}CT}\right)
+
\frac{e^{-2\pi^{2}CT}}{\sqrt{2\pi^{3}CT}}
\right].
\end{equation}
Using the relation \eqref{matching_green_sigma} between the UV and IR Green's functions, we obtain the imaginary part of the retarded UV Green's function as
\begin{equation}
\operatorname{Im} G^{R}_{\mathrm{IR}}
=
3 \,  \omega
\left[
\operatorname{erf}\!\left(\sqrt{2\pi^{2}CT}\right)
\left(1+\frac{1}{4\pi^{2}CT}\right)
+
\frac{e^{-2\pi^{2}CT}}{\sqrt{2\pi^{3}CT}}
\right].
\end{equation}
Then, the quantum-corrected DC conductivity is given by the Kubo formula
\begin{equation}\label{eq:conductivity_final_linear_axion}
    \sigma_{\mathrm{DC}} = \lim_{\omega \to 0} \frac{1}{\omega}\operatorname{Im} G^R_{\mathrm{UV}} = 3 \left[\operatorname{erf}\!\left(\sqrt{2\pi^{2}CT} \right) \left(1+\frac{1}{4\pi^{2}CT}\right) + \frac{e^{-2\pi^{2}CT}}{\sqrt{2\pi^{3}CT}} \right].
\end{equation}
In the two extreme regimes of $CT$, the conductivity behaves as
\begin{align}
\sigma_{\mathrm{DC}}
=
\begin{cases}
\displaystyle
\frac{3\sqrt{2}}{\sqrt{\pi^3 CT}} \left(1+\frac{2\pi^2}{3} CT+ \ldots \right),
& \qquad CT \ll 1,
\\[4mm]
\displaystyle
3 \left(1+\frac{1}{4 \pi ^2 CT}+ O(e^{-CT}) \right),
& \qquad CT \gg 1.
\end{cases}
\end{align}

\begin{figure}[t]
    \centering
    \includegraphics[width=0.6\textwidth]{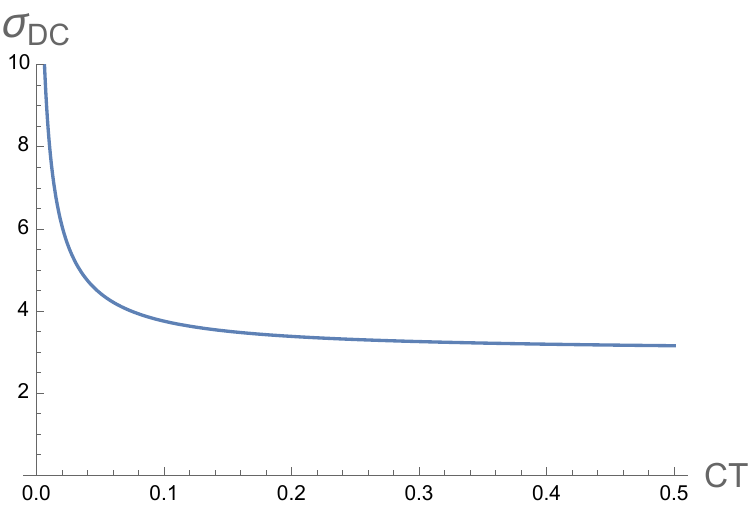}
    \caption{Behavior of DC conductivity $\sigma_{\mathrm{DC}}$, given in Equation \eqref{eq:conductivity_final_linear_axion}, as a function of $CT$.}
    \label{conduct_linear_axion}
\end{figure}

Figure \ref{conduct_linear_axion} shows the behavior of \(\sigma_{\mathrm{DC}}\) as a function of the dimensionless parameter \(CT\). The conductivity becomes significantly large for small \(CT\) and clearly diverges as \(CT \rightarrow 0\). As \(CT\) increases, \(\sigma_{\mathrm{DC}}\) decreases monotonically, and eventually approaches a constant asymptotic value in the regime \(CT \gg 1\). The low temperature divergence mirrors the behavior previously observed for $\eta$. The pattern strongly hints at a universal low-temperature divergence of transport coefficients in holographic field theories, provided the Schwarzian quantum corrections are appropriately incorporated. We will address this universality in Section \ref{sec:universality}; for the moment we note that the monotonic behavior of the conductivity is observed for $\Delta=1$.

\section{Nonzero gauge field background and chemical potential} \label{Sec:ChemicalPotential}
Having clarified most of the conceptual and computational details in a simple model with momentum relaxation, we now turn to a more complete one that incorporates a non-trivial chemical potential. This more generic model allows us to interpolate among various particular cases and understand the role of each ingredient more clearly. We follow the setup of \cite{Andrade:2013gsa} to calculate the shear viscosity and conductivity with a nonzero gauge field background:
\begin{equation}
ds^2 = -f(r)\,dt^2 + \frac{dr^2}{f(r)} + r^2 4\pi V_{2} \delta_{ab}\,dx^a dx^b,
\quad
A = A_t(r)\,dt,
\quad
\psi_I = \sqrt{4\pi V_{2}}\,\alpha_{Ia}\,x^a,
\end{equation}
with 
\begin{subequations}
\begin{align}
f &= r^2 - \frac{\alpha^2}{2(d-2)} - \frac{m_0}{r^{d-2}} + \frac{(d-2)\mu^2}{2(d-1)} \frac{r_{+}^{2(d-2)}}{r^{2(d-2)}},\\
A_t &= \mu \left(1 - \frac{r_{+}^{d-2}}{r^{d-2}}\right),
\end{align}
\end{subequations}
where $r_{+}$ is the location of the outer horizon.
\begin{equation}
m_0 = r_{+}^d \left( 1 + \frac{d-2}{2(d-1)} \frac{\mu^2}{r_{+}^2} - \frac{1}{2(d-2)} \frac{\alpha^2}{r_{+}^2} \right),
\end{equation}
which is a parameter proportional to the energy density of the brane.
The temperature of the brane is given by
\begin{equation}
T = \frac{f'(r_{+})}{4\pi}
= \frac{1}{4\pi}\left( d r_{+} - \frac{\alpha^2}{2r_{+}} - \frac{(d-2)^2 \mu^2}{2(d-1)r_{+}} \right).
\end{equation}
Near extremality, we obtain the small temperature expansions of different black hole parameters as follows
\begin{subequations}
    \begin{align}
        r_{+} &= r_{0} + \frac{2\pi}{3} T + O \left(T^{2} \right),\\
        r_{-} &= r_{0} - \frac{2\pi}{3 \left(\frac{12\, r_{0}^{2}}{\alpha^{2}} - 1 \right)} T + O \left(T^{2} \right),\\
        m_{0} &= r_{0} \left(4 r_{0}^{2} - \alpha^{2} \right) + \frac{2\pi^{2} \left(12 r_{0}^{2} - \alpha^{2} \right)}{9 r_{0}} T^{2} + O\left(T^{3} \right).
    \end{align}
\end{subequations}
The small temperature expansion is carried out by keeping the chemical potential fixed at $\mu = \sqrt{12 r_{0}^2 - 2\alpha^{2}}$. The above expressions suggest that the entropy density behaves as
\begin{equation}
    s=\frac{r_+^2}{4 G_N}
    =
    s_0 \left( 1+\frac{4 \pi}{3} \frac{T}{r_0} +O(T^2) \right),
\end{equation}
where $s_0$ denotes the extremal entropy density. As before, the linear-in-temperature behavior of the entropy, compared to the quadratic temperature scaling of the mass, implies a breakdown of semiclassical black hole thermodynamics. The energy scale below which this breakdown occurs is given by
\begin{equation}\label{eq:breakingscale_full}
\varepsilon_{\text{gap}} \equiv 4\pi^2
\left(\frac{\partial s}{\partial T} \right)^{-1}
= \frac{12 \pi G}{r_{0}}.
\end{equation}

In the following, we will introduce different sets of perturbations on the above background to compute the holographic shear viscosity and DC conductivity.

\subsection{Shear viscosity}
\label{sec:Non-zero gauge field, eta/s}
To compute the holographic shear viscosity, one can obtain the equation for the shear perturbation directly from equation (4.22) of \cite{Andrade:2013gsa} by setting $k=0$:
\begin{equation}
\left(r^{2}f\,H_{xy}'\right)'
+\left(\frac{r^{2}\omega^{2}}{f}-\alpha^{2}\right)H_{xy}=0
\end{equation}

\subsubsection*{Inner region}
We perform coordinate transformation $\zeta=\frac{r-r_0}{\omega}$, and take $\omega\to 0$. This results in
\begin{equation} \label{eq:hxy_chargedbh_inner_eom}
\zeta^2 h''+2\zeta h'
+\left[
\frac{4r_0^4}{(12r_0^2-\alpha^2)^2}\frac{1}{\zeta^2}
-\frac{2\alpha^2}{12r_0^2-\alpha^2}
\right]h=0,
\end{equation}
where $h\left(\zeta \right) \equiv H_{xy} \left(\zeta \right)$.
This is a Bessel-type equation whose general solution is 
\begin{equation}
h(\zeta)=\sqrt{\frac{\lambda}{\zeta}}
\left[A_{\rm in}\,H_\nu^{(1)}\!\left(\frac{\lambda}{\zeta}\right)
+A_{\rm out}\,H_\nu^{(2)}\!\left(\frac{\lambda}{\zeta}\right)
\right],
\end{equation}
where $H_{\nu}^{(1)}$, $H_{\nu}^{(1)}$ are Hankel functions of order $\nu$,
\begin{equation}
    \lambda=\frac{2r_0^2}{12r_0^2-\alpha^2},\quad\nu
=\frac12\sqrt{\frac{12r_0^2+7\alpha^2}{12r_0^2-\alpha^2}}.
\end{equation}
As the appropriate boundary condition we keep the ingoing solution,
\begin{equation}
    h_{\rm in}(\zeta) = \frac{i \pi \lambda^{-\frac{1}{2} + \nu}}{2^{\nu}\, \Gamma(\nu)}\sqrt{\frac{\lambda}{\zeta}}\,
H_\nu^{(1)}\!\left(\frac{\lambda}{\zeta}\right).
\end{equation}
We have chosen a specific normalization to simplify some of the following expressions. As a consistency check, when $\alpha \to 0$ and $\alpha \to \sqrt{6} r_{0}$, one obtains $\nu=\frac{1}{2}$ and $\nu = \frac{3}{2}$, respectively. The solution $h_{\mathrm{in}} \left(\zeta \right)$ reduces to 
\begin{equation}
    h_{\rm in}(\zeta) =
        \begin{cases}
            e^{\frac{i}{6\zeta}},\quad &\alpha = 0\\
            \left(\zeta - \frac{i}{3} \right) e^{\frac{i}{3 \zeta}},\quad &\alpha = \sqrt{6} r_{0}
        \end{cases}
\end{equation}
which agrees with the RN-AdS$_{4}$ result in \cite{Edalati:2009bi} as well as Equation \eqref{near_region} for the neutral black hole case.

In the overlapping region, $\zeta\to \infty$, with the help of the asymptotic behavior of Hankel function
\begin{equation}
    H_\nu^{(1)}(x)\sim
-\frac{i\Gamma(\nu)}{\pi}\left(\frac{2}{x}\right)^\nu
+\frac{1+i\cot(\pi\nu)}{\Gamma(\nu+1)}
\left(\frac{x}{2}\right)^\nu
+\cdots,\qquad x\to0,
\end{equation}
we notice that the solution behaves as
\begin{equation} \label{eq:hxy_chargedbh_inner_sol_larger}
    h_{\rm in}(\zeta)
\sim
\zeta^{-\frac12 + \nu} + \mathcal{B}\,\zeta^{-\frac12 - \nu}
+\cdots,\quad \mathcal{B} = \frac{i \pi \lambda^{2\nu} \left(1 + i \cot \pi \nu \right)}{2^{2\nu}\, \Gamma(\nu) \Gamma(1 + \nu)}.
\end{equation}

Returning to the original $r$ coordinate, the asymptotic behavior of the inner solution near the AdS$_{2}$ conformal boundary takes the following form
\begin{equation}
    \left. h_{\text{in}} \left(r \right) \right|_{r \to \infty} = \left(\frac{r - r_{0}}{\omega} \right)^{-\frac{1}{2} + \nu} \left(1 + O\left(\frac{1}{r-r_{0}}\right) \right) + \mathcal{B} \left( \frac{r - r_{0}}{\omega} \right)^{-\frac{1}{2} - \nu} \left(1 + O\left(\frac{1}{r-r_{0}}\right) \right).
\end{equation}
Therefore, according to the standard rules of gauge/gravity duality, we obtain the IR retarded Green's function to be given by
\begin{equation} \label{eq:GIR_charged_bh}
    \mathcal{G}_{\mathrm{IR}} \left(\omega\right) = \mathcal{B} \omega^{2\nu}.
\end{equation}

\subsubsection*{Outer region}

In the region of spacetime situated far from the horizon, we switch to the dimensionless radial coordinate $u = \frac{r - r_{0}}{r_{0}}$. In the $\omega \to 0$ limit, the leading order equation in this outer region is  easily obtained to be 
\begin{equation}
    h''(u) + \frac{2 \left(4 r_{0}^2 \left(u \left(u + 3 \right) + 3 \right) - \alpha^{2} \right)}{u \left(2 r_{0}^2 \left(u \left(u + 4 \right) + 6 \right) - \alpha^{2} \right)} h'(u) - \frac{2 \alpha^{2}}{u^2 \left(2 r_{0}^{2} \left(u \left(u + 4 \right) + 6 \right) -\alpha^{2} \right)} h(u) = 0.
\end{equation}
Let us check the leading terms in this equation in the limit $u \to 0$, we find that in this limit the equation is well-approximated by
\begin{equation}
    h''(u)  + \frac{2}{u} h'(u) - \frac{2\alpha^{2}}{u^{2} \left(12 r_{0}^{2} - \alpha^{2} \right)} h(u) = 0.
\end{equation}
The last equation happens to be identical to the inner region Equation \eqref{eq:hxy_chargedbh_inner_eom} when $\zeta \to \infty$, a fact already observed in \cite{Faulkner:2009wj}. Therefore, it is convenient to write the two linearly independent solutions $\mathsf{h}_{\pm}$ in the outer region using the two linearly independent terms in Equation \eqref{eq:hxy_chargedbh_inner_sol_larger}, i.e. $\mathsf{h}_{\pm}$ \emph{are chosen to satisfy the boundary condition}
\begin{equation}
    \mathsf{h}^{(0)}_{\pm} (r) \approx \left(r - r_{0} \right)^{-\frac{1}{2} \pm \nu} + \cdots,\quad r \rightarrow r_{0},
\end{equation}
where ellipses denote terms that vanish as $r \to r_{0}$. The matching with the near-horizon region then becomes trivial, and the leading order solution $h^{(0)}_{\text{out}} (u)$ in the outer region could be written as
\begin{equation}
    h^{(0)}_{\text{out}} (u) = \mathsf{h}_{+}^{(0)}(u) + \mathcal{G}_{\text{IR}} \left(\omega\right)\, \mathsf{h}^{(0)}_{-}(u),
\end{equation}
where $\mathcal{G}_{\text{IR}}\left(\omega\right)$ is the IR Green's function given in Equation \eqref{eq:GIR_charged_bh}. For higher orders in $\omega$, we may expand the two linearly independent solutions as
\begin{equation}
    \mathsf{h}_{\pm}(u) = \mathsf{h}^{(0)}_{\pm}(u) + \omega\, \mathsf{h}^{(1)}_{\pm}(u) + \omega^{2}\, \mathsf{h}^{(2)}_{\pm}(u) + \cdots,
\end{equation}
and obtain $\mathsf{h}^{(n)}_{\pm}(u)$, $n \geq 1$ by perturbatively solving the outer region equation. In general, we conclude that
\begin{equation}
\label{eq:boundary condition for hout}
    h_{\text{out}} (u) = \mathsf{h}_{+}(u) + \mathcal{G}_{\text{IR}} \left(\omega\right)\, \mathsf{h}_{-}(u).
\end{equation}

Near the AdS$_{4}$ conformal boundary $\left( r \to \infty \right)$, the functions admit the usual expansions
\begin{equation}
    \left. \mathsf{h}_{\pm}(r) \right|_{r \to \infty} = \left(a_{\pm}^{(0)} + \omega\,a^{(1)}_{\pm} + O\left(\omega^{2} \right) \right) r^{ \Delta-3} + \left(b_{\pm}^{(0)} + \omega\,b^{(1)}_{\pm} + O\left(\omega^{2} \right) \right) r^{-\Delta}.
\end{equation}
The retarded Green's function is easily determined by the real-time AdS/CFT prescription \cite{Faulkner:2009wj}
\begin{equation}
    G_{\mathrm{UV}} = \textcolor{red}{-3} \frac{b^{(0)}_{+} + \mathcal{G}_{\mathrm{IR}} \, b^{(0)}_{-}}{a^{(0)}_{+} + \mathcal{G}_{\mathrm{IR}} a^{(0)}_{-} }.
\end{equation}
Here we have neglected the higher order terms in $\omega$. The unknown coefficients $a^{(n)}_{\pm}$ and $b^{(n)}_{\pm}$ can only be obtained by numerically solving the full outer region equation order by order.
\begin{figure}[t]
    \centering
    \includegraphics[width=0.8\linewidth]{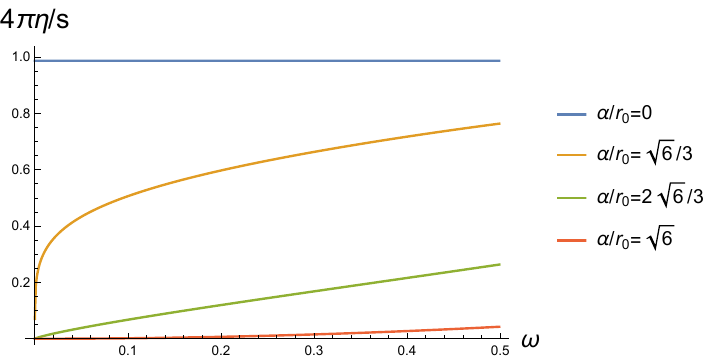}
    \caption{The behavior of $\eta/s$ with $\omega$ for different values of the relaxation parameter $\alpha$.}
    \label{fig:shear-viscosity-classical-generic-alpha}
\end{figure}

The semiclassical results for $\frac{4\pi \eta}{s}$ as a function of the frequency $\omega$ is shown in Figure \ref{fig:shear-viscosity-classical-generic-alpha}. As expected for any nonzero $\alpha$, $\eta$ smoothly approaches zero when $\omega \to 0$.

\subsection*{Quantum Corrections}

As we saw earlier, in order to compute the quantum-corrected transport coefficients, we need the Schwarzian Green's function, since $\mathcal{G}_{IR}$ must be replaced by the Schwarzian-corrected retarded correlator. The conformal dimension of the Schwarzian correlator is given by
\begin{equation}
    \Delta=\frac{1}{2}+\nu, \qquad 
    \nu=\frac12\sqrt{\frac{12r_0^2+7\alpha^2}{12r_0^2-\alpha^2}}.
\end{equation}
Unlike the previous section, $\Delta$ now depends on the ratio $\alpha/r_0$. For the allowed range of $\alpha/r_0$, one has $1 \leq \Delta \leq 2$, where $\Delta=1$ and $\Delta=2$ correspond to the ordinary Reissner--Nordström case and the case discussed in Section \ref{sec:quantum_Green}, respectively.

It is not possible to obtain a closed-form expression for the Green's function \eqref{sw_correlator_k} for generic $\Delta$. Therefore, we evaluate the integral appearing in \eqref{sw_correlator_k} numerically. It is worth emphasizing that, in the limit $\omega \rightarrow 0$, the result becomes independent of $\omega$. Consequently, the retarded IR Green's function contains only an imaginary part, which is completely fixed by the Wightman function, as shown in \eqref{eq:retarded_green}. With this, the quantum-corrected retarded UV Green's function takes the form
\begin{equation}
    G^R_{\mathrm{UV}} = -3
    \frac{b^{(0)}_{+}+b^{(0)}_{-}G^R_{\Delta}}
    {a^{(0)}_{+}+a^{(0)}_{-}G^R_{\Delta}}.
\end{equation}
From this expression, we can extract the imaginary part in order to compute $\eta$. The quantum-corrected entropy density can also be computed straightforwardly using \eqref{eq:corrected_s}, with $C=V_2/\varepsilon_{\text{gap}}$, where $\varepsilon_{\text{gap}}$ is given in \eqref{eq:breakingscale_full}.

\begin{figure}[t]
    \centering
    \includegraphics[width=0.8\textwidth]{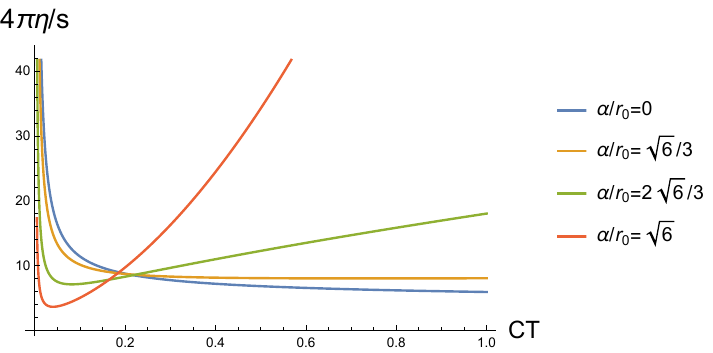}
    \caption{Quantum-corrected $\eta/s$ for different values of relaxation parameter $\alpha$ for $V_2=3/4$ and $r_0=1$.}
    \label{eta_by_s_full_geometry}
\end{figure}

Figure \ref{eta_by_s_full_geometry} shows the behavior of the quantum-corrected ratio $\eta/s$ for different values of the momentum-relaxation parameter $\alpha/r_0$. In the absence of momentum relaxation $\left(\frac{\alpha}{r_{0}} = 0 \right)$, the quantity \(4\pi \eta/s \) approaches unity in the limit \(CT \to \infty\). This observation is in agreement with previous findings reported in \cite{Cremonini:2025yqe, Kanargias:2025vul}. In contrast, when momentum relaxation is present, \(4\pi \eta/s\) diverges as \(CT \to \infty\). For any nonzero 
$\alpha$, the ratio exhibits a minimum between these two extreme limits, and the location of the minima shifts toward smaller values of \(CT\) as \(\alpha/r_0\) increases.
%

\subsection{DC conductivity}
In order to compute the holographic conductivity, we introduce the following linearized perturbations
\begin{equation}
\delta A_x = e^{-i\omega t} a_x(r),
\qquad
\delta g_{tx} = e^{-i\omega t} r^2 H_{tx}(r).
\qquad
\delta \psi_1 = e^{-i\omega t}\,\alpha^{-1}\chi(r).
\end{equation}
Following \cite{Andrade:2013gsa}, let us consider the field redefinition
\begin{equation}
\phi = \omega^{-1} r^{d-1} f \chi'.
\end{equation}
The linearized equations of motion 
 for $\{a_x,\,H_{tx},\,\chi\}$ can be diagonalized by two master fields $\Phi_{\pm}$ which are defined as \cite{Andrade:2013gsa}
\begin{equation}
\phi = r^{d-2}\left(c_{+}\Phi_{+} + c_{-}\Phi_{-}\right),
\qquad
a_x = -i\left(\Phi_{+} + \Phi_{-}\right),
\end{equation}
with
\begin{equation}
c_{\pm} = \frac{1}{2\mu r_0^{d-2}}
\left[
d m_0 \pm \left(d^2 m_0^2 + 4 r_0^{2(d-2)} \mu^2 \alpha^2\right)^{1/2}
\right].
\end{equation}
The equations for the master fields decouple 
\begin{equation} \label{eq:master_eq_conductivity}
r^{5-d}\left(f r^{d-3}\Phi_{\pm}'\right)'
+
\left(
\frac{r^2\omega^2}{f}
-
\frac{(d-2)^2 \mu^2 r_0^{2(d-2)}}{r^{2(d-2)}}
+
c_{\pm}(d-2)\mu \frac{r_0^{d-2}}{r^{d-2}}
\right)\Phi_{\pm}
= 0.
\end{equation}
For simplicity, let us take $d=3$. As we are interested in the transport near extremality, we also take the extremal limit $T=0$.
\begin{equation}
\label{eq:Phi_+}
r^{2}\bigl(f_{\rm ext}\Phi_+'\bigr)'
+\left[
\frac{r^{2}\omega^{2}}{f_{\rm ext}}
+(12r_0^2-2\alpha^2)\!\left(\frac{r_0}{r}-\frac{r_0^2}{r^2}\right)
\right]\Phi_+=0,
\end{equation}
\begin{equation}
\label{eq:Phi_-}
r^{2}\bigl(f_{\rm ext}\Phi_-'\bigr)'
+\left[
\frac{r^{2}\omega^{2}}{f_{\rm ext}}
-\frac{\alpha^2 r_0}{r}
-(12r_0^2-2\alpha^2)\frac{r_0^2}{r^2}
\right]\Phi_-=0,
\end{equation}
where
\begin{equation}
f_{\rm ext}(r)=\frac{(r-r_0)^2}{r^2}\left(r^2+2r_0r+3r_0^2-\frac{\alpha^2}{2}\right).
\end{equation}
The chemical potential parameter $\mu$ has been eliminated by the $T=0$ condition.

\subsubsection*{Inner region}

In the near-horizon region, we perform the coordinate transformation $\zeta=\frac{r-r_0}{\omega}$ and take the small frequency limit $\omega\to0$. The equations \eqref{eq:master_eq_conductivity} then reduce to
\begin{subequations}
    \begin{align}
        \zeta^2 \Phi_+''+2\zeta \Phi_+'+\frac{4r_0^4}{(12r_0^2-\alpha^2)^2}\frac{1}{\zeta^2}\Phi_{+} &= 0, \label{eq:Inner Phi_p} \\
        \zeta^2 \Phi_-''+2\zeta \Phi_-'+\left(\frac{4r_0^4}{(12r_0^2-\alpha^2)^2}\frac{1}{\zeta^2}-2\right)\Phi_{-} &= 0, \label{eq:Inner Phi_m}
    \end{align}
\end{subequations}
which have solutions
\begin{subequations}
    \begin{align}
        \Phi_+(\zeta) &= A_+^{\rm in}e^{\,i\lambda/\zeta}+A_+^{\rm out}e^{-\,i\lambda/\zeta}\\
        \Phi_-(\zeta) &= A_-^{\rm in}\Bigl(1+i\frac{\zeta}{\lambda}\Bigr)e^{\,i\lambda/\zeta}
        + A_-^{\rm out}\Bigl(1-i\frac{\zeta}{\lambda}\Bigr)e^{-\,i\lambda/\zeta}.
    \end{align}
\end{subequations}
Imposing the ingoing boundary condition at the horizon, we set $A_+^{\rm out},A_-^{\rm out}=0$. Expanding the solutions around $\zeta\to\infty$, we see that the two master fields have different conformal dimensions: $\Delta(\Phi_+)=1$, while $\Delta(\Phi_-)=2$. The ingoing solutions have the forms
\begin{subequations}
    \begin{align}
        \Phi_{+} (r) &= A_+^{\rm in}\left[(1+\dots)+\frac{\mathcal{G}_{IR,+}}{r-r_0}(1+\dots)\right],\\
        \Phi_{-} (r) &= A_-^{\rm in}\frac{i}{\lambda\omega}\left[(r-r_0)(1+\dots)+\frac{\mathcal{G}_{IR,-}}{(r-r_0)^2}(1+\dots)\right],
    \end{align}
\end{subequations}
with the IR Green's functions being given by
\begin{equation}
    \mathcal{G}_{IR,+}=i \lambda \omega,\qquad\mathcal{G}_{IR,-}=\frac{i\lambda^3\omega^3}{3}.
\end{equation}

\subsection*{Outer region}

Similar to Section \ref{sec:Non-zero gauge field, eta/s}, in the outer region, we can safely take $\omega\to0$ in equations \eqref{eq:Phi_+} and \eqref{eq:Phi_-}. After performing the coordinate transformation to $u$ and taking the limit of $u\to0$, the equations take the form 
\begin{subequations}
    \begin{align}
        \Phi_{+}''(u) + \frac{2}{u}\Phi_{+}'(u) &= 0,\\
        \Phi_{-}''(u) + \frac{2}{u}\Phi_{-}'(u) - \frac{2}{u^2}\Phi_{-}(u) &= 0.
    \end{align}
\end{subequations}
The solutions of these equations also have the same asymptotic behavior at $u\to0$ as that of \eqref{eq:Inner Phi_p} and \eqref{eq:Inner Phi_m}. Similar to \eqref{eq:boundary condition for hout}, we can impose the following boundary condition for $\Phi_{\pm}$ in the outer region in terms of the two independent solutions at $u\to0$,
\begin{equation}
    \Phi_{\pm}(u)=\Phi_{1,\pm}(u)+\mathcal{G}_{IR,\pm}(\omega)\Phi_{2,\pm}(u),
\end{equation}
where, differently from \eqref{eq:boundary condition for hout}, we use $\pm$ to denote the solutions of $\Phi_\pm$, the indices $\{1,2\}$ are used to denote the two independent solutions with different powers as $u \to 0$. Around the AdS$_{4}$ boundary, both $\Phi_+$ and $\Phi_-$ have the same asymptotic structure
\begin{equation}
    \Phi_{\pm}(r)=A(1+\dots)+\frac{B}{r}(1+\dots).
\end{equation}
Expanding the two independent solutions around the AdS$_{4}$ conformal boundary, $r\to\infty$, we obtain
\begin{equation}
        \left. \Phi_{\{1,2\},\pm}(r) \right|_{r \to \infty} = \left(a_{\{1,2\},\pm}^{(0)} + \omega\,a^{(1)}_{\{1,2\},\pm} + O\left(\omega^{2} \right) \right) + \left(b_{\{1,2\},\pm}^{(0)} + \omega\,b^{(1)}_{\{1,2\},\pm} + O\left(\omega^{2} \right) \right) r^{-1}.
\end{equation}
Noticing $a_x = -i\left(\Phi_{+} + \Phi_{-}\right)$, the UV retarded Green's function of the gauge field perturbation $a_x$ is
\begin{equation}\label{eq:UV_IR_conductivity}
       G_{\mathrm{UV}} =-
    \frac{
    \sum_{\sigma=\pm} C_\sigma
    \left[
    b^{(0)}_{1,\sigma}
    +
    \mathcal{G}_{\mathrm{IR},\sigma}(\omega)b^{(0)}_{2,\sigma}
    \right]
    }{
    \sum_{\sigma=\pm} C_\sigma
    \left[
    a^{(0)}_{1,\sigma}
    +
    \mathcal{G}_{\mathrm{IR},\sigma}(\omega)a^{(0)}_{2,\sigma}
    \right]
    },
\end{equation}
where $C_\sigma$ is taken to satisfy
\begin{equation}
    c_-C_-(a_{1,-}+a_{2,-}\mathcal G_{IR,-})+c_+C_+(a_{1,+}+a_{2,+}\mathcal G_{IR,+})=0,    
\end{equation}
to turn off the boundary source of the scalar fields. We have neglected the terms of higher order in $\omega$.\\



\begin{figure}[t]
    \centering
    \includegraphics[width=0.6\linewidth]{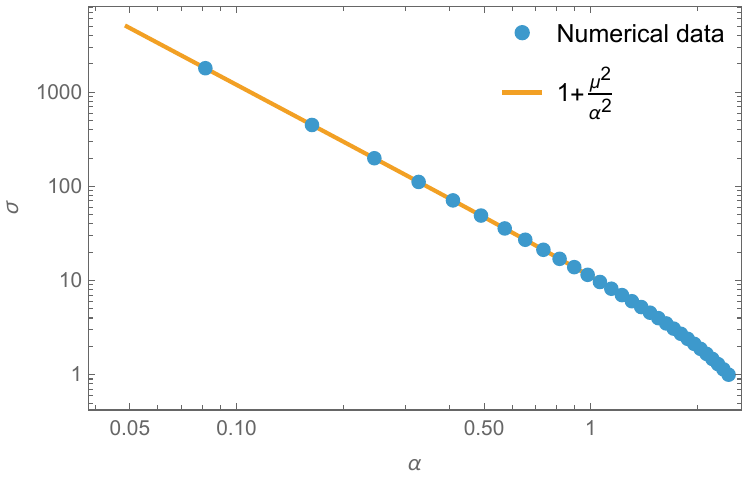}
    \caption{Semiclassical holographic conductivity, $\sigma_{\mathrm{DC}}$ with $r_0=1$, $\alpha\in(0,\sqrt{6}r_0]$.}
    \label{fig:Conductivity_NonzeroGaugeField}
\end{figure}

After numerically solving for the connection coefficients $\lbrace a_{\lbrace 1, 2 \rbrace, \sigma}, b_{\lbrace 1, 2 \rbrace, \sigma}^{(0)}, C_{\sigma} \rbrace$, the holographic DC conductivity is obtained by using the Kubo formula
\begin{equation} \label{eq:Kubo_formula_sigma} \sigma=-\lim_{\omega\to0}\frac{1}{\omega}\rm Im\,G_{\mathrm{UV}}.\end{equation}
The numerical result is shown in Figure \ref{fig:Conductivity_NonzeroGaugeField}. It reproduces the result of \cite{Andrade:2013gsa}
\begin{equation}
    \sigma_{\rm DC}=r_0^{d-3}\left(1+(d-2)^2\frac{\mu^2}{\alpha^2} \right),
\end{equation}
for $d=3$. At the right end, $\dfrac{\alpha}{r_{0}} = \sqrt{6}$, our result agrees with \eqref{eq:sigam_zero_gauge_field} which corresponds to $\mu=0$.

By contrast, the work \cite{Edalati:2009bi} showed that in the absence of momentum relaxation, i.e., $\alpha=0$, which corresponds to the left end in Figure \ref{fig:Conductivity_NonzeroGaugeField}, the conductivity should be zero. This disagreement between \cite{Edalati:2009bi} and \cite{Andrade:2013gsa} originates from the fact that the limits $\omega\to 0$ and $\alpha\to 0$ do not commute. In Figure \ref{fig:Conductivity_NonzeroGaugeField}, we have taken the limit $\omega\to0$ first, which is the choice of \cite{Andrade:2013gsa}. For the left endpoint, $\alpha=0$, if instead we take the limit $\alpha\to0$ first, our procedure can also reproduce the result of \cite{Edalati:2009bi}.

In Appendix \ref{App:A-W-Limits}, we show how these two limits do not commute and explain how our numerical framework reproduces such expression in a particular order of limits.

\subsection*{Quantum Corrections}


In order to compute the quantum corrections to holographic conductivity, we replace the semiclassical IR retarded Green's functions $\mathcal{G}_{\mathrm{IR},\pm}$ in \eqref{eq:UV_IR_conductivity} with the quantum-corrected retarded Green's functions $G^R_{\Delta=1,2}$. As before, we first take the limit $\omega \rightarrow 0$. To this end, we expand \eqref{eq:UV_IR_conductivity} for small $\omega$, obtaining
\begin{equation}
    G_{\mathrm{UV}}= \frac{-1}{1-\frac{c_-}{c_+}} \left[ \left( \frac{b_{1,-}}{a_{1,-}} -\frac{c_-}{c_+}   \frac{b_{1,+}}{a_{1,+}} \right)  -\frac{c_-}{c_+} \left( \frac{a_{1,+}b_{2,+}-a_{2,+}b_{1,+}}{a_{1,+}^2} \right) \mathcal{G}_{\mathrm{IR},+} +O(\omega^2)   \right].
\end{equation}
The first term inside the square brackets is a constant, independent of $\omega$, and is purely real. Therefore, it does not contribute to the real part of the conductivity. The higher-order terms also do not contribute in the $\omega \rightarrow 0$ limit. Consequently, only the second term contributes to the Kubo formula, and we obtain 
\begin{equation} \label{conduct_0}
    \sigma \approx -\frac{c_-}{c_+-c_-} \left( \frac{a_{1,+}b_{2,+}-a_{2,+}b_{1,+}}{a_{1,+}^2} \right) \frac{\mathrm{Im}\,\mathcal{G}_{\mathrm{IR},+}}{\omega}.
\end{equation}
The expression \eqref{conduct_0} is valid only when $\dfrac{\alpha}{r_{0}} < \sqrt{6}$. At the extreme value, $\alpha = \sqrt{6}\, r_{0}$, $c_-$ blows up; the consistent choice here is to set $C_-=0$. This leads to the following expression
\begin{equation}
    \sigma \approx \left( \frac{a_{1,+}b_{2,+}-a_{2,+}b_{1,+}}{a_{1,+}^2} \right) \frac{\mathrm{Im}\,\mathcal{G}_{\mathrm{IR},+}}{\omega}.
\end{equation}
Our numerical results indeed reproduce
\begin{equation}
    \sigma = \frac{3\, \mathrm{Im}\,\mathcal{G}_{\mathrm{IR},+}}{\omega},
\end{equation}
for $\alpha/r_0=\sqrt{6}$, in agreement with the result presented in Section \ref{sec:classical_Green}.
\begin{figure}[t]
    \centering
    \includegraphics[width=0.8\textwidth]{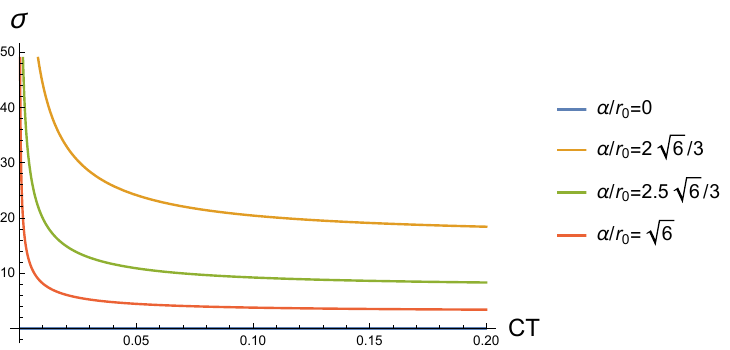}
    \caption{The quantum-corrected conductivity $\sigma$ for different values of relaxation parameter $\alpha$. Other parameters: $V_2=3/4, r_0=1$. The monotonic behavior of $\sigma$ is correlated with  $\Delta=1$ for all these values of $\alpha/r_0$.}
    \label{sigma_quantum_full_geometry}
\end{figure}

We now replace $\mathcal{G}_{\mathrm{IR},+}$ with the quantum-corrected retarded IR Green's function for $\Delta=1$. The corresponding Wightman function is given in \eqref{eqn:Green_delta1}, from which the retarded Green's function can be obtained using \eqref{eq:reln_Wight_Retarded}. Substituting these results into \eqref{conduct_0}, we obtain
\begin{align}
    \sigma  \approx  \frac{c_-(a_{2,+}b_{1,+}-a_{1,+}b_{2,+})}{a_{1,+}^2 (c_+-c_-)} 
\left(
\operatorname{erf}\!\left(\sqrt{2\pi^{2}CT}\right)
\left(1+\frac{1}{4\pi^{2}CT}\right)
+
\frac{e^{-2\pi^{2}CT}}{\sqrt{2\pi^{3}CT}}
\right).
\end{align}
This expression has the same functional form as \eqref{eq:conductivity_final_linear_axion}, derived in Section \ref{sec:quantum_Green}, differing only by an overall normalization factor. In Figure \ref{sigma_quantum_full_geometry}, we present the quantum-corrected DC conductivity as a function of $CT$ for different values of the momentum relaxation parameter $\alpha/r_0$.  For any nonzero $\alpha$, the conductivity approaches a constant value for large values of $CT$. A constant asymptotic value is always determined by the $\Delta=1$ Green's function. As mentioned earlier, for $\alpha=0$, $c_-=0$ so conductivity is zero and lies along the $x$ axis. 
%


\section{A universal feature of quantum corrections} \label{sec:universality}

Given the evidence accumulated in the explicit treatment of holographic transport in various models, we now venture to argue that the behavior observed in the quantum-corrected transport coefficients is, in fact, universal. Generically, transport coefficients determined via the quantum-corrected retarded Green's functions have a minimum.

The Kubo formula relates the retarded UV Green's functions of various fluctuations to the corresponding transport coefficients. For a (near-)extremal black hole, an emergent long AdS$_2$ throat develops in the near-horizon region. One can define a retarded Green's function at the boundary of this AdS$_2$ region, which we refer to as the IR Green's function. The relation between the UV and IR Green's functions can be obtained through the matching procedure discussed in the previous sections. Let us consider the following generic form of the UV Green's function:
\begin{equation}
\label{eq:reln}
G_{\mathrm{UV}}
=
\frac{a+b\,\mathcal{G}_{IR}}
     {c+d\,\mathcal{G}_{IR}} \, .
\end{equation}
Locality of the dual quantum field theory requires $\mathcal{G}_{IR}$ to scale with a positive power of $\omega$. Therefore, in the hydrodynamic limit $\omega \to 0$, Eq.~\eqref{eq:reln} can be expanded as
\begin{equation}
G_{\mathrm{UV}}
\approx
\frac{a}{c}
+
\frac{bc-ad}{c^2}\,\mathcal{G}_{IR}
+\cdots .
\end{equation}
To compute the quantum corrections to the transport coefficients, one must replace the classical IR retarded Green's function $\mathcal{G}_{IR}$ by the retarded Schwarzian Green's function, which can be extracted from the corresponding Schwarzian Wightman function.

For generic values of $\Delta$, no closed-form analytic expression for the Wightman function is known. However, it admits the following series representation (see Appendix~\ref{App:GreenDelta} for details):
\begin{equation}
\label{G_any_delta1}
\begin{aligned}
G_{\Delta}
&=
\frac{2^{\Delta+\frac{1}{2}}
      \Gamma(\Delta)^2
      (CT)^{\Delta-\frac{1}{2}}}
     {\pi^{\frac{3}{2}}
      C^{2\Delta-1}
      \Gamma(2\Delta)}
\,e^{-2\pi^2CT}
\\
&\quad\times
\Bigg[
\Gamma(\Delta+1)\,
{}_1F_1\!\left(
\Delta+1;\frac{3}{2};2\pi^2CT
\right)
+\frac{a_1\Gamma(\Delta)}
       {2CT}\,
{}_1F_1\!\left(
\Delta;\frac{3}{2};2\pi^2CT
\right)
\\
&\qquad
+\frac{a_2\Gamma(\Delta-1)}
       {(2CT)^2}\,
{}_1F_1\!\left(
\Delta-1;\frac{3}{2};2\pi^2CT
\right)
+\ldots
\Bigg] ,
\end{aligned}
\end{equation}
where the coefficients $a_i$ are given in appendix~\ref{App:GreenDelta}. The corresponding retarded Green's function, following the KMS relations, is
\begin{equation}
G_{\Delta}^{R}
=
\frac{i C \omega}{2CT}\,
G_{\Delta}.
\end{equation}
Let us now explain that  $G_{\Delta}^{R}$ diverges in the limit $CT\to0$. The details of the expansion of  $G_\Delta$  as indicated in Equation \eqref{G_any_delta1} are worked out in Appendix~\ref{App:GreenDelta}.  The crucial point is that the expansion takes the form $G_\Delta^R \sim (CT)^{\Delta-\frac{3}{2}}\left(1+\frac{1}{CT}+\frac{1}{(CT)^2}+\cdots\right).$ Therefore, it always diverges. However, for a general value of $\Delta$, it is not straightforward to determine the leading divergent term. For $\Delta<3/2$, the first term is divergent; for $3/2<\Delta<5/2$, it is the second term in the expansion that diverges, etc. Consequently, the quantum-corrected IR Green's function $\mathcal{G}_{IR}$ also diverges in this limit. It then follows from eq.~\eqref{eq:reln} that the UV Green's function becomes singular as $CT\to0$. In various particular cases, the divergence is of the form $1/\sqrt{CT}$ reflecting that the leading divergence arises from $\Delta=1$. 

On the other hand, using the large-$CT$ behavior of Eq.~\eqref{G_any_delta1}, given in Eq.~\eqref{G_delta_CT_large}, one finds that
\begin{equation}
G_{\mathrm{UV}}
\sim
(CT)^{2\Delta-2},
\qquad
CT\to\infty .
\end{equation}
For $\Delta>1$, the UV Green's function therefore also diverges in the large-$CT$ limit. Therefore, the transport coefficient becomes large both in the $CT\to0$ and $CT\to\infty$ limits. Since it is not identically infinite for all values of $CT$, this suggests that its behavior cannot be monotonic and that there must exist at least one minimum at an intermediate value of $CT$. 

An exception to the above behavior arises in the precise case where $\Delta=1$. In this case the behavior for large $CT$ is constant and a completely monotonic behavior of the corresponding transport is possible. This is precisely the case for the electrical conductivity in Section \ref{sec:quantum_Green} given in Figure \ref{conduct_linear_axion}. A similar situation occurs for the  conductivity presented in Figure \ref{sigma_quantum_full_geometry}.



\section{Summary and conclusions} \label{sec:conclusions}

Momentum relaxation is a key ingredient toward holographic constructions of realistic transport; it is crucial to obtain non-divergent DC conductivities. In this manuscript, we have studied the low-temperature quantum corrections in a specific holographic model of momentum relaxation. We expect, based on the universality within various models  of momentum relaxation mechanisms described in \cite{Hartnoll:2016tri}, that  our results are indicative of generic behavior present in  various models of momentum relaxation.

Our principal results are presented in Sections \ref{sec:quantum_Green}  and \ref{Sec:ChemicalPotential}. In Section \ref{sec:quantum_Green} we computed the quantum-corrected shear viscosity and electrical conductivity in a minimal model of momentum relaxation. We found that for $\Delta > 1$, the transport coefficients develop a characteristic minimum at a temperature $T_{\rm min}$, below which they increase as $1/\sqrt{CT}$. In Section \ref{Sec:ChemicalPotential}  we generalized the analysis to a more realistic holographic model including both momentum relaxation and finite chemical potential. Remarkably, the same qualitative behavior persists: both the viscosity and the electrical conductivity exhibit an enhancement as the temperature is lowered into the region $CT\ll 1$.

Taken together with previous analyses of viscosity \cite{PandoZayas:2025snm,Gouteraux:2025exs,Kanargias:2025vul} and conductivity \cite{Sabyasachi-Leo-Jingchao}, these results suggest a universal role for Schwarzian quantum fluctuations. Rather than simply renormalizing transport coefficients, they qualitatively alter their infrared behavior, replacing the classical low-temperature scaling by a non-monotonic temperature dependence characterized by a minimum at $T_{\rm min}$.  In Section \ref{sec:universality} we provided evidence that this phenomenon is generic for transport associated with operators of arbitrary effective conformal dimension $\Delta >1$. The special case $\Delta =1$ emerges as the unique exception, for which the transport coefficient remains monotonic and approaches a constant at larger temperatures.

Several interesting directions naturally follow from our analysis. A particularly important problem is to develop the complete hydrodynamic description of these quantum-corrected systems, including the associated spectrum of quasinormal modes. Progress in this direction has recently been made in \cite{Nian:2025oei}, but many aspects of the interplay between quantum near-horizon dynamics and hydrodynamics remain to be understood.

Our results point toward a deeper connection between quantum black holes and quantum transport. For example, the relation between  $\eta/s$ and the graviton absorption cross section suggests that the low-temperature enhancement of viscosity found here has a direct interpretation in terms of quantum properties of the near-extremal black hole. Although the notion of viscosity is modified in the presence of momentum relaxation, it is natural to expect that analogous gravitational observables encode the corresponding quantum corrections to transport. Developing a precise holographic dictionary between quantum dynamics in the near-horizon region and the infrared transport properties of the dual field theory remains an important challenge.

 \acknowledgments{
 We are thankful to  Blaise Gout\'eraux, Xiao-Long Liu, Jun Nian, Cl\'ement Supiot, Joaqu\'in Turiaci, Cong-Yuan Yue for insightful discussions. This work was partially supported by the U.S. Department of Energy under Grant No. DE-SC0007859 and by the National Natural Science Foundation of China (NSFC) under Grant No. 12247103. }

\appendix

\section{Green's function for arbitrary $\Delta$}\label{App:GreenDelta}

We now attempt to obtain a closed-form expression for the Schwarzian Green’s function for generic values of $\Delta$. To this end, we rewrite eq.~\eqref{sw_correlator_k}, the momentum-space Green’s function, as
\begin{equation}\label{sw_correlator_k2}
\begin{aligned}
G_{\Delta}(\omega)
&=
\frac{e^{S_0}(2C)}{Z(T)\,4\pi^{4}(2C)^{2\Delta}\Gamma(2\Delta)}
\int_{0}^{\infty} dk \;
2k \,
e^{-\frac{k^{2}}{2CT}}
\sinh(2\pi k)
\\
&\quad \times
\sinh\!\left(2\pi\sqrt{k^{2}+2C\omega}\right)
\prod_{\sigma_1,\sigma_2=\pm1}
\Gamma\!\left(
\Delta
+i\sigma_1\sqrt{k^{2}+2C\omega}
+i\sigma_2 k
\right).
\end{aligned}
\end{equation}
To derive an analytic expression for the Green’s function, we focus on the regime $k \gg C\omega$ where 
\begin{align}
&\prod_{\sigma_1,\sigma_2=\pm1}
\Gamma\!\left(
\Delta+i\sigma_1\sqrt{k^2+2C\omega}
+i\sigma_2 k
\right)
\nonumber\\[2mm]
&=
\Gamma(\Delta)^2
\Gamma(\Delta+2ik)\Gamma(\Delta-2ik)
\Bigg(
1
+\frac{iC\omega}{k}
\Big[
\psi(\Delta+2ik)-\psi(\Delta-2ik)
\Big]
+O\!\left[\left(\frac{C\omega}{k}\right)^2\right]
\Bigg).
\end{align}
It is worth mentioning that the $O(\omega)$ or higher terms do not contribute to the transport coefficient as, by definition, they require taking the limit $\omega\to 0$. To obtain the retarded Green's function from the Wightman Green's function one multiplies by a factor of $\omega$, whereas the Kubo formula contains a compensating factor of $1/\omega$. As a result, the $O(\omega)$ and higher order corrections vanish in the limit $\omega \to 0$, provided that $\omega$ is the smallest scale in the problem. Therefore, we keep only the leading $O(1)$ term, which in the large-$k$ limit can be approximated as
\begin{equation}
\prod_{\sigma_1,\sigma_2=\pm1}
\Gamma\!\left(
\Delta+i\sigma_1\sqrt{k^2+2C\omega}
+i\sigma_2 k
\right)
\approx 2\pi \, \Gamma(\Delta)^2 \, (2k)^{2\Delta-1} e^{-2\pi k}
\left(1+\frac{a_1}{k^2}+\frac{a_2}{k^4}+\frac{a_3}{k^6} + \ldots \right).
\end{equation}
The coefficients are given by
\begin{align}
a_1 &= \frac{\Delta(\Delta-1)(2\Delta-1)}{24}, \\
a_2 &= \frac{\Delta(\Delta-1)(\Delta-2)(2\Delta-1)(2\Delta-3)(5\Delta+1)}{5760}, \\
a_3 &= \frac{\Delta(\Delta-1)(\Delta-2)(\Delta-3)(2\Delta-1)(2\Delta-3)(2\Delta-5)(35\Delta^2+21\Delta+4)}{2903040}.
\end{align}

\noindent
Using $\sinh(2\pi k)\simeq (\frac{1}{2}) e^{2\pi k}$ for $k\gg 1$, Eq.~\eqref{sw_correlator_k2} reduces to
\begin{align}
G_{\Delta}(\omega)
&=
\frac{e^{S_0} \Gamma(\Delta)^2}{2\pi^{3} Z(T) \, C^{2\Delta-1}\Gamma(2\Delta)}
\int_{0}^{\infty} dk \;
k^{2\Delta} \,
e^{-\frac{k^{2}}{2CT}}
\sinh(2\pi k)
\left(1+\frac{a_1}{ k^2} +\frac{a_2}{k^4} +\frac{a_3}{k^6}+ \ldots \right) \nonumber \\
&=
\frac{e^{S_0} \Gamma(\Delta)^2 2^{\Delta}}{\pi^{2} Z(T) \, C^{2\Delta-1}\Gamma(2\Delta)}
(CT)^{\Delta+1}
\Bigg[
\Gamma(\Delta+1) \, _1F_1\!\left(\Delta+1;\frac{3}{2};2 \pi ^2 C T\right)
\nonumber \\
&\quad
+\frac{a_1 \Gamma(\Delta)}{2 C T}  \, _1F_1\!\left(\Delta;\frac{3}{2};2 \pi ^2 C T\right)
+\frac{a_2 \Gamma(\Delta-1)}{(2 C T)^2}  \, _1F_1\!\left(\Delta-1;\frac{3}{2};2 \pi ^2 C T\right)
+ \ldots
\Bigg].
\end{align}
Substituting the partition function from \eqref{eq:partition_function1}, we finally obtain
\begin{equation}\label{G_any_delta}
\begin{aligned}
G_{\mathrm{IR}}^{\Delta}(\omega)
&=
\frac{2^{\Delta+\frac{1}{2}} \Gamma(\Delta)^2 \, (CT)^{\Delta-\frac{1}{2}} }{\pi^{\frac{3}{2}} C^{2\Delta-1} \Gamma(2 \Delta)} \,
e^{-2\pi^2 C T}
\\
&\quad \times
\Bigg[
\Gamma(\Delta+1) \, _1F_1\!\left(\Delta+1;\frac{3}{2};2 \pi ^2 C T\right)
+\frac{a_1 \Gamma(\Delta)}{2 C T}  \, _1F_1\!\left(\Delta;\frac{3}{2};2 \pi ^2 C T\right)
\\
&\qquad
+\frac{a_2 \Gamma(\Delta-1)}{(2 C T)^2}  \, _1F_1\!\left(\Delta-1;\frac{3}{2};2 \pi ^2 C T\right)
+ \ldots
\Bigg].
\end{aligned}
\end{equation}
This is the desired IR Green’s function. We now consider its behavior in two limiting regimes.

\subsubsection*{$CT \gg 1$ regime}

In this limit, we obtain
\begin{equation}\label{G_delta_CT_large}
G_{\mathrm{IR}}^{\Delta}(\omega)
\approx
\frac{\pi^{2\Delta-\frac{3}{2}}\,\Gamma(\Delta)}{C^{2\Delta-1}\Gamma\!\left(\Delta+\frac{1}{2}\right)}
(CT)^{2\Delta-1}
\left(
1
+
\frac{\Delta(2\Delta-1)}{4\pi^2 CT}
+\ldots
\right).
\end{equation}

\subsubsection*{$CT \ll 1$ regime}

This limit is more subtle. The subleading terms in the large-$k$ expansion become increasingly important, and in general it is difficult to obtain a closed-form expression for arbitrary $\Delta$. However, for certain special values of $\Delta$, only a finite number of coefficients $a_i$ are nonzero, allowing for an exact closed-form expression. We list a few such cases below:
\begin{align}
G_{\mathrm{IR}}^{\Delta=1}
&\sim \frac{1}{C}\sqrt{\frac{8}{\pi^3}} \sqrt{CT}
\left( 1+\frac{2\pi^2}{3} CT-\frac{2\pi^4}{15} (CT)^2 +\ldots \right), \\
G_{\mathrm{IR}}^{\Delta=\frac{3}{2}}
&\sim \frac{1}{128 C^2}
\left(1+\frac{3}{8} CT+\frac{\pi^2}{2} (CT)^2+\ldots \right), \\
G_{\mathrm{IR}}^{\Delta=2}
&\sim \frac{1}{C^3}\sqrt{\frac{1}{72\pi^3}} \sqrt{CT}
\left( 1+\frac{2(\pi^2+24)}{3} CT-\frac{2\pi^2(\pi^2-240)}{15} (CT)^2 +\ldots \right), \\
G_{\mathrm{IR}}^{\Delta=\frac{5}{2}}
&\sim \frac{147}{2048 C^4}
\left(1+\frac{45}{1024} CT+\frac{15(6+\pi^2)}{256} (CT)^2+\ldots \right).
\end{align}

\begin{figure}[h]
\centering
    \includegraphics[width=0.45\linewidth]{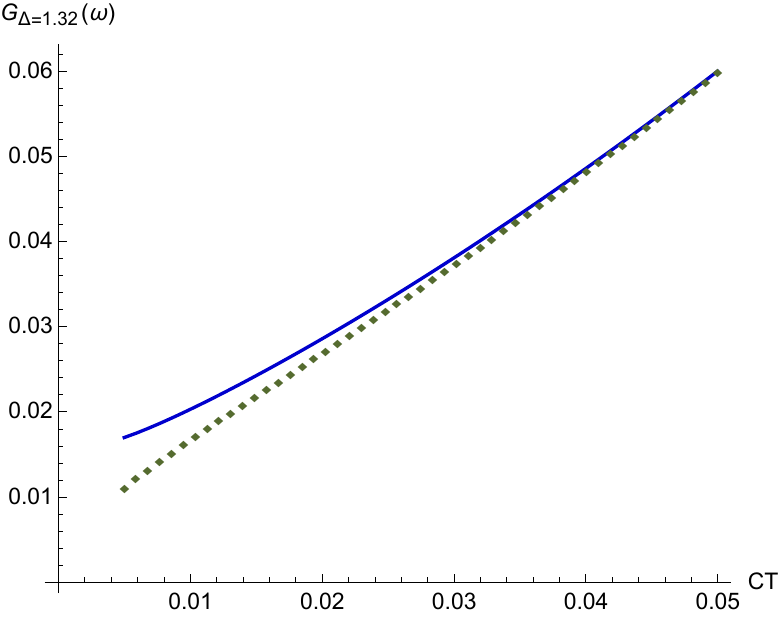}
    \hfill
    \includegraphics[width=0.45\linewidth]{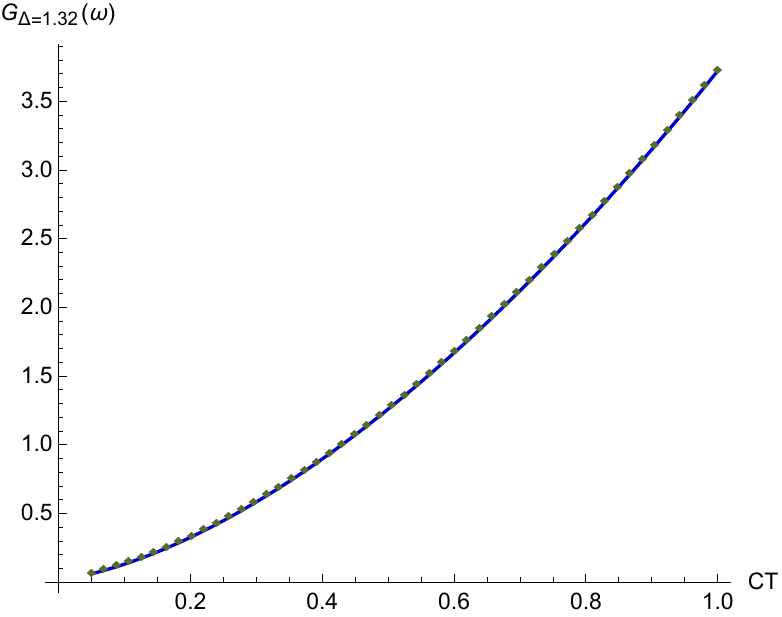}
    \caption{The figures show a comparison between the exact Green's function of eq.~\eqref{sw_correlator_k2}, obtained from numerical integration (shown as discrete dots), and its analytic approximate form \eqref{G_any_delta} (shown as a continuous blue curve), plotted as a function of $CT$. The plots correspond to $\Delta = 1.32$.}
    \label{comparison_green1}
\end{figure}

For generic $\Delta$, all terms in the expansion are non-vanishing. In the limit $CT \ll 1$, contributions that are increasingly suppressed by higher powers of $k$ in the large-$k$ expansion become progressively important, and consequently the truncated expression \eqref{G_any_delta} deviates from the exact result, as illustrated in Figure~\ref{comparison_green1} for $\Delta=1.32$.

\begin{figure}[t]
    \centering
    \includegraphics[width=0.6\textwidth]{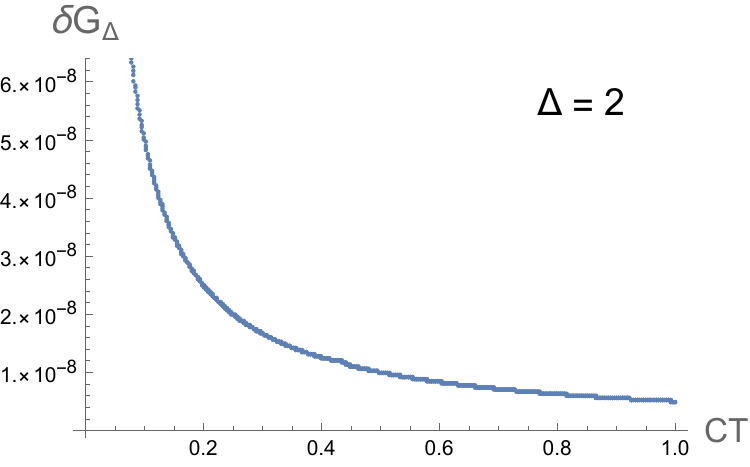}
    \caption{Figure shows the behavior of relative error, $\delta G_{\Delta}=\left|\frac{G^{\rm Numeric}-G^{\rm Analytic}}{G^{\rm Numeric}}\right|,$ as a function of $CT$ for $\Delta=2$ and $\omega=10^{-8}$.}
    \label{comparison_Green}
\end{figure}

As for generic $\Delta$, we do not have an analytic expression, and therefore the results presented in the main text are obtained numerically. To demonstrate the agreement between the numerical and analytic results, in Figure \ref{comparison_Green} we plot the relative error, $\delta G_{\Delta}=\left|\frac{G^{\rm Numeric}-G^{\rm Analytic}}{G^{\rm Numeric}}\right|,$
as a function of $CT$ for $\Delta=2$. Although the relative error increases with $CT$, it remains of order $10^{-8}$ throughout the range considered. This is of the same order as $\omega$, the smallest scale in the problem, which is ultimately taken to zero.

%
%
%
%

%
%

\newpage

\section{Normalization of boundary coefficients}
\label{app:normalization-boundary-coefficients}

In this appendix we clarify the normalization of the retarded Green's function. To that end, we derive the factors that must multiply the ratios of the two asymptotic coefficients obtained from the large-\(u\) expansions, where
\(u=r/r_0\).  We avoid absorbing those factors into the definitions of the boundary
coefficients.

We use the following action with an explicit gravitational coupling and an
explicit Maxwell normalization,
\begin{equation}
\label{eq:A1}
S
=
\frac{1}{16\pi G}
\left[
\int_M d^4x\,\sqrt{-g}
\left(
R-2\Lambda
-\frac12\sum_{I=1}^{2}(\partial\psi_I)^2
-\frac{1}{4g_F^2}F_{\mu\nu}F^{\mu\nu}
\right)
-2\int_{\partial M}d^3x\,\sqrt{-\gamma}\,K
\right]
+S_{\rm ct}.
\end{equation}
Here the factor \(1/(16\pi G)\) multiplies the Maxwell term as well.  The background
metric is
\begin{equation}
\label{eq:A2}
ds^2
=
-f(r)dt^2
+
\frac{dr^2}{f(r)}
+
\frac{r^2}{L^2}\,4\pi V_2
\left(dx_1^2+dx_2^2\right).
\end{equation}
The determinant and inverse radial metric component are
\begin{equation}
\label{eq:A4}
\sqrt{-g}
=
\frac{4\pi V_2 r^2}{L^2},
\qquad
g^{rr}=f(r).
\end{equation}
Near the AdS$_4$ boundary,
\begin{equation}
\label{eq:A5}
f(r)
=
\frac{r^2}{L^2}
+\cdots .
\end{equation}
We use the real-time prescription
\begin{equation}
\label{eq:A6}
S_{\rm os}^{(2)}
=
\int\frac{d\omega}{2\pi}\,
\varphi_0(-\omega)\,
\mathcal F(\omega)\,
\varphi_0(\omega)
\quad
\Longrightarrow
\quad
G^R(\omega)=-2\mathcal F(\omega).
\end{equation}
With this convention, the Kubo formulae are written as
\begin{equation}
\label{eq:A7}
\eta
=
-\lim_{\omega\to0}
\frac{1}{\omega}
\operatorname{Im}G^R_{x_1x_2,x_1x_2}(\omega),
\qquad
\sigma_{\rm DC}
=
-\lim_{\omega\to0}
\frac{1}{\omega}
\operatorname{Im}G^R_{x_1x_1}(\omega).
\end{equation}

\subsection{Metric fluctuation}

We define the shear fluctuation by
\begin{equation}
\label{eq:A8}
\delta g_{x_1x_2}
=
g_{x_1x_1}\,h(t,r)
=
\frac{r^2}{L^2}\,4\pi V_2\,h(t,r),
\qquad
h(t,r)=e^{-i\omega t}h(r).
\end{equation}
The fluctuation \(h\) is dimensionless.  Expanding the Einstein-Hilbert action
together with the Gibbons-Hawking term to quadratic order gives
\begin{equation}
\label{eq:A9}
S_h^{(2)}
=
-\frac{1}{32\pi G}
\int d^4x\,\sqrt{-g}\,
g^{MN}\partial_M h\,\partial_N h
+S^{(2)}_{h,{\rm local}} .
\end{equation}
The terms collected in \(S^{(2)}_{h,{\rm local}}\) are local in the boundary fields and
can only change contact terms.  The radial derivative part is
\begin{equation}
\label{eq:A10}
S_{h,{\rm rad}}^{(2)}
=
-\frac{1}{32\pi G}
\int dr\,dt\,d^2x\,
\sqrt{-g}\,g^{rr}
h'(-\omega,r)h'(\omega,r).
\end{equation}
Using the metric data,
\begin{equation}
\label{eq:A11}
\sqrt{-g}\,g^{rr}
=
\frac{4\pi V_2 r^2}{L^2}\,f(r).
\end{equation}
Dividing by the spatial volume factor \(4\pi V_2\), the radial part of the action
density is
\begin{equation}
\label{eq:A12}
s_{h,{\rm rad}}^{(2)}
=
-\frac{1}{32\pi G}
\int dr\,\frac{d\omega}{2\pi}\,
\frac{r^2}{L^2}f(r)
h'(-\omega,r)h'(\omega,r).
\end{equation}
After integrating by parts and using the fluctuation equation, the UV boundary term is
\begin{equation}
\label{eq:A13}
s_{h,{\rm os}}^{(2)}
=
-\frac{1}{32\pi G}
\int\frac{d\omega}{2\pi}
\left[
\frac{r^2}{L^2}f(r)
h(-\omega,r)\partial_r h(\omega,r)
\right]_{r\to\infty}
+s_{h,{\rm ct}}^{(2)} .
\end{equation}
In the physical radial coordinate \(r\), the large-\(r\) expansion is
\begin{equation}
\label{eq:A14}
h(r)
=
h^{(0)}
+
\frac{h_r^{(3)}}{r^3}
+\cdots .
\end{equation}
Therefore
\begin{equation}
\label{eq:A15}
\partial_r h(r)
=
-\frac{3h_r^{(3)}}{r^4}
+\cdots .
\end{equation}
Using \(f(r)\sim r^2/L^2\), one finds
\begin{equation}
\label{eq:A16}
\frac{r^2}{L^2}f(r)\partial_r h(r)
=
-\frac{3h_r^{(3)}}{L^4}
+\cdots .
\end{equation}
Thus
\begin{equation}
\label{eq:A17}
s_{h,{\rm os}}^{(2)}
=
\frac{3}{32\pi G L^4}
\int\frac{d\omega}{2\pi}\,
h^{(0)}(-\omega)h_r^{(3)}(\omega)
+s_{h,{\rm ct}}^{(2)} .
\end{equation}
Equivalently,
\begin{equation}
\label{eq:A22}
\mathcal F_h(\omega)
=
\frac{3}{32\pi G L^4}
\frac{h_u^{(3)}(\omega)}{h^{(0)}(\omega)}
+\hbox{contact terms}.
\end{equation}
Using \eqref{eq:A6}, the retarded correlator is
\begin{equation}
\label{eq:A23}
G^R_{x_1x_2,x_1x_2}(\omega)
=
-\frac{3}{16\pi G L^4}
\frac{h_u^{(3)}(\omega)}{h^{(0)}(\omega)}
+\hbox{contact terms}.
\end{equation}
Thus the normalization factor multiplying the VEV-to-source ratio in the
shear channel is
\begin{equation}
\label{eq:A24}
\mathcal N_\eta
=
\frac{3}{16\pi G L^4}.
\end{equation}

\subsection{Gauge-field fluctuation}

We next consider the transverse Maxwell fluctuation
\begin{equation}
\label{eq:A25}
A_{x_1}(t,r)
=
e^{-i\omega t}a_{x_1}(r),
\qquad
A_r=0 .
\end{equation}
The Maxwell part of the action is
\begin{equation}
\label{eq:A26}
S_A
=
-\frac{1}{16\pi G}
\frac{1}{4g_F^2}
\int d^4x\,\sqrt{-g}\,
F_{\mu\nu}F^{\mu\nu}.
\end{equation}
For the fluctuation \eqref{eq:A25},
\begin{equation}
\label{eq:A27}
F_{rx_1}
=
\partial_r a_{x_1},
\qquad
F_{tx_1}
=
-i\omega a_{x_1}.
\end{equation}
The radial derivative part of the quadratic Maxwell action is
\begin{equation}
\label{eq:A28}
S_{A,{\rm rad}}^{(2)}
=
-\frac{1}{32\pi G g_F^2}
\int dr\,dt\,d^2x\,
\sqrt{-g}\,
g^{rr}g^{x_1x_1}
a_{x_1}'(-\omega,r)a_{x_1}'(\omega,r).
\end{equation}
For the metric \eqref{eq:A2},
\begin{equation}
\label{eq:A29}
g^{x_1x_1}
=
\frac{L^2}{4\pi V_2 r^2},
\end{equation}
therefore, 
\begin{equation}
\label{eq:A30}
\sqrt{-g}\,g^{rr}g^{x_1x_1}
=
f(r).
\end{equation}
The radial Maxwell action becomes
\begin{equation}
\label{eq:A31}
S_{A,{\rm rad}}^{(2)}
=
-\frac{1}{32\pi G g_F^2}
\int dr\,\frac{d\omega}{2\pi}\,
f(r)
a_{x_1}'(-\omega,r)a_{x_1}'(\omega,r).
\end{equation}
After integrating by parts and using the Maxwell equation, the UV boundary term is
\begin{equation}
\label{eq:A32}
S_{A,{\rm os}}^{(2)}
=
-\frac{1}{32\pi G g_F^2}
\int\frac{d\omega}{2\pi}
\left[
f(r)
a_{x_1}(-\omega,r)\partial_r a_{x_1}(\omega,r)
\right]_{r\to\infty}
+S_{A,{\rm ct}}^{(2)} .
\end{equation}
In the physical radial coordinate \(r\), the large-\(r\) expansion is
\begin{equation}
\label{eq:A33}
a_{x_1}(r)
=
a^{(0)}
+
\frac{a_r^{(1)}}{r}
+\cdots .
\end{equation}
Therefore
\begin{equation}
\label{eq:A34}
\partial_r a_{x_1}(r)
=
-\frac{a_r^{(1)}}{r^2}
+\cdots .
\end{equation}
Using \(f(r)\sim r^2/L^2\), we obtain
\begin{equation}
\label{eq:A35}
f(r)\partial_r a_{x_1}(r)
=
-\frac{a_r^{(1)}}{L^2}
+\cdots .
\end{equation}
Thus
\begin{equation}
\label{eq:A36}
S_{A,{\rm os}}^{(2)}
=
\frac{1}{32\pi G g_F^2 L^2}
\int\frac{d\omega}{2\pi}\,
a^{(0)}(-\omega)a_r^{(1)}(\omega)
+S_{A,{\rm ct}}^{(2)} .
\end{equation}
Equivalently,
\begin{equation}
\label{eq:A41}
\mathcal F_A(\omega)
=
\frac{1}{32\pi G g_F^2 L^2}
\frac{a_u^{(1)}(\omega)}{a^{(0)}(\omega)}
+\hbox{contact terms}.
\end{equation}
Using \eqref{eq:A6}, the retarded current correlator is
\begin{equation}
\label{eq:A42}
G^R_{x_1x_1}(\omega)
=
-\frac{1}{16\pi G g_F^2 L^2}
\frac{a_u^{(1)}(\omega)}{a^{(0)}(\omega)}
+\hbox{contact terms}.
\end{equation}
Thus the normalization factor multiplying the VEV-to-source ratio in the
conductivity channel is
\begin{equation}
\label{eq:A43}
\mathcal N_\sigma
=
\frac{1}{16\pi G g_F^2 L^2}.
\end{equation}

\section{The order of limits in the semiclassical computation of holographic conductivity} \label{App:A-W-Limits}

In this section we revisit the various orders of limits implicit in the original computation of conductivity in the model of momentum relaxation as reported  in \cite{Andrade:2013gsa}. Recall that the results reported in  \cite{Andrade:2013gsa} take the following analytic form 
\begin{equation}
\label{eq:C1}
    \sigma_{\rm DC}=r_0^{d-3}\left(1+(d-2)^2\frac{\mu^2}{\alpha^2} \right).
\end{equation}
To obtain the above result, the authors of \cite{Andrade:2013gsa} used a radially conserved quantity $\Pi$ in their Equation (3.14). From Equation (3.13) of \cite{Andrade:2013gsa}, it is observed that $\Pi$ is only conserved when $\omega=0$. Therefore, when we take the limit $\alpha \to 0$ in \eqref{eq:C1}, we have taken $\omega \to 0$ first. However, in \cite{Edalati:2009bi}, the model has no momentum relaxation in the first place, so it is equivalent to taking $\alpha \to 0$ first, followed by the limit $\omega \to 0$ when using the Kubo formula. 

Let us explicitly verify that these two limits do not commute. We perform a ``pseudo-analytical" calculation using the procedure in Section \ref{Sec:ChemicalPotential}. Although we do not have analytical control over the connection coefficients $a_{\{1,2\},\pm}, b_{\{1,2\},\pm}$, the numerical results show that as $\alpha\to0$, $a_{1,+} \propto \alpha^2$. All other connection coefficients go to some finite but nonzero values which are not important to control the qualitative behavior of $\sigma_{\mathrm{DC}}$. Let us assume $a_{1,+}\to k_{1,+}\alpha^2$ as $\alpha \to 0$, and then expand $\rm Im\, G_{\mathrm{UV}}$ in the following two ways:
\begin{itemize}
    \item Powers of $\omega$ first
    \begin{equation}
        {\rm Im} \, G_{\mathrm{UV}}=\left(-\frac{a_{2,+}b_{1,+}}{72\, k_{1,+}\alpha^2}+\mathcal O(\alpha^0)\right)\omega+\mathcal O(\omega^3)
    \end{equation}
    \item Powers of $\alpha$ first
    \begin{equation}
        {\rm Im} \, G_{\mathrm{UV}}=\left(-\frac{a_{2,-}b_{1,-}+a_{1,-}b_{2,-}}{648\, a_{1,-}^2}\omega^3+\mathcal O(\omega^4)\right)+\mathcal O(\alpha^2)
    \end{equation}
\end{itemize}
Upon using the Kubo formula \eqref{eq:Kubo_formula_sigma}, the two choices above yield $\sigma_{\mathrm{DC}} \to \alpha^{-2}$, and $\sigma_{\mathrm{DC}} \to 0$, respectively.







\bibliography{references_green}
\bibliographystyle{JHEP.bst}

\end{document}